\documentclass[%
 amsmath,amssymb,
 aps,prl, %Journal of the APS family, PRL paper
 reprint,%
]{revtex4-1}

\usepackage{graphicx}% Include figure files
\usepackage{dcolumn}% Align table columns on decimal point
\usepackage{bm}% bold math
\usepackage[usenames, dvipsnames]{color}
\usepackage{hyperref}% add hypertext capabilities
\hypersetup{
  colorlinks   = true,    % Colours links instead of boxes
  urlcolor     = blue,    % Colour for external hyperlinks
  linkcolor    = blue,    % Colour of internal links
  citecolor    = blue      % Colour of citations
}
\usepackage{multirow}
\usepackage{array}
\usepackage{booktabs}
\newcolumntype{M}[1]{>{\centering\arraybackslash}m{#1}}
\usepackage{textgreek}
\usepackage{nicefrac} %package for fraction display (Table I)
\setcitestyle{numbers,square,citesep={,\kern-.24em}} %remove space after comma in citation
\usepackage[referable]{threeparttablex}
\begin{document}

\title{Compression of Exact Wavefunctions with Restricted Boltzmann Machine Auto-Encoders}

\author{Anderson D. S. Duraes}
\email{anderson.duraes@nd.edu}
%\begingroup
%\hypersetup{linkcolor=black}
%\endgroup
\affiliation{%
 Department of Chemistry and Biochemistry, University of Notre Dame, 251 Nieuwland Science Hall, Notre Dame, IN 46556, USA 
}%

%\date{\today}
\date{July 2018}
\begin{abstract}
    Virtually, every ab-initio electronic structure method (Coupled Cluster, DMRG, etc.) can be viewed as an algorithm to compress the ground-state wavefunction. This compression is usually obtained by exploiting some physical structure of the wavefunction, which leads to issues when the system changes and that structure is lost. Compressions which are efficient near equilibrium (coupled cluster) or in 1-D systems (DMRG) often fail catastrophically elsewhere. To overcome these issues, we seek a scheme that compresses wavefunctions without any supervised physical information. In this manuscript, we introduce a scheme to compress molecular wavefunctions using a model for high dimensional functions from machine learning: a restricted Boltzmann machine (RBM). We show that, while maintaining chemical accuracy, the RBM can significantly compress the exact wavefunction.
\end{abstract}

\maketitle

\section{Introduction}

\indent In his Nobel lecture, Kohn stressed the problem of storing an accurate many-body wavefunction~$(\Phi)$ for a large system on a classical computer.~\cite{Kohn1999} For simple and direct model chemistries, like the full configuration interaction~(FCI) method, the storage problem is essentially the main stumbling block to exact improvable results.~\cite{Bauschlicher1989,Roos1972,Olsen1990,Shepard1990,Szabo1989,Mayer2003,Helgaker2000} The FCI method employs a linear combination of all the possible Slater determinants ($\Psi_{n}$'s) in order to span the exact wavefunction~$(\Phi_{\text{FCI}})$:~\cite{Bauschlicher1989,Szabo1989,Mayer2003,Helgaker2000} 
\begin{align}
    \Phi_{\text{FCI}}=\sum_{n=0}\;c_{n}\Psi_{n}\text{.}
    \label{fci_disc_eq}
\end{align}
\indent However, depending on the quantity of electrons and atomic orbitals of a system, the full set of electronic $\Psi_{n}$'s---and, consequently, the number of bits---are simply too numerous to manipulate on a classical machine; forbidding any FCI calculation for even modestly sized molecules.~\cite{Szabo1989,Bauschlicher1989,Taylor2013,Olsen1990,Helgaker2000}

\indent In order to face this storage problem, many authors have tried to compress ~$\Phi_{\text{FCI}}$.~\cite{Bauschlicher1989,Taylor2013,McClean2015,Knowles2015,Marti2010,Buenker1974,Zhang2016,Bytautas2004,Bytautas2005,Zimmerman2017,Bytautas2011,Alcoba2014,Alcoba2016,Huron1973,Evangelisti1983,Harrison1991,Roth2009,Holmes2016,Tubman2016,Schriber2017,Booth2009,Cleland2010,Booth2014,Petruzielo2012,Ten-no2013,Greer1995,Tong2000,Coe2012,Kelly2014,Holmes2016a,Ohtsuka2017} These compression algorithms are usually based on physical insights into the structure of the exact wavefunction or based on mathematical insights into approximate solutions of the ground state problem. These compressions exploit the fact that only a small fraction of the $\Psi_{n}$'s (Eq.~\ref{fci_disc_eq}) usually contribute to an accurate ground state wavefunction.~\cite{Ivanic2001,Bytautas2009}

\indent For instance, the selected CI plus perturbation theory correction (SCI+PT) algorithms~\cite{Huron1973,Evangelisti1983,Harrison1991,Roth2009,Holmes2016,Tubman2016,Schriber2017}---such as the Heat-Bath CI~(HBCI)~\cite{Holmes2016}---implement deterministic constraints to select configuration expansions which significantly contribute to an accurate ground-state energy. Alternatively, Monte Carlo algorithms~\cite{Booth2009,Cleland2010,Booth2014,Petruzielo2012,Ten-no2013,Greer1995,Tong2000,Coe2012,Kelly2014,Holmes2016a,Ohtsuka2017}---such as the FCI Quantum Monte Carlo (FCIQMC)~\cite{Booth2009,Cleland2010,Booth2014}---implement stochastic constraints to select configuration expansions. Both methods are able to treat larger CI spaces than a naive approach. On the other hand, these [deterministic/stochastic] constraints are somewhat arbitrary, generating a systematic source of error for the estimated FCI calculations.~\cite{Azar2015,Zimmerman2017}

\indent We are instead curious about compressing the Slater determinants without any specific physical or mathematical structure, using a neural network to achieve a non-linear map. To do this, we apply the Restricted Boltzmann Machine (RBM). 

\indent RBM~\cite{Smolensky1986,Freund1992,Hinton2012,Hinton2006,Chen2017,LeRoux2008,Montufar2011,Carleo2017} is classified as an unsupervised learning algorithm and its structure consists of two layers: one layer having the visible units and the other, the binary hidden units.~\cite{Smolensky1986,Freund1992,Hinton2012,Hinton2006,Chen2017,LeRoux2008,Montufar2011} The visible units process the input data, and the hidden units designs the compression of the input.~\cite{Smolensky1986,Freund1992,Hinton2012,Hinton2006,Chen2017,LeRoux2008,Montufar2011} The bridge between the two layers---visible units to hidden units---is established by parameters that connect both units in a process denominated as encoding.~\cite{Smolensky1986,Freund1992,Hinton2012,Hinton2006,Bengio2012,Kingma2013,Chen2017,LeRoux2008,Montufar2011} The reverse process, known as decoding, uses the binary hidden units---with the same parameters in the encoding---to recover the uncompressed (original) input data.~\cite{Smolensky1986,Freund1992,Hinton2012,Hinton2006,Bengio2012,Kingma2013,Chen2017,LeRoux2008,Montufar2011}

\indent In addition, RBM has found successful application to compress images~\cite{Hinton2006a,Torralba2008,Cheng2016,Lokare2015}, to model data~\cite{Salakhutdinov2007,Larochelle2008,Freund1992,Xing2012,Fiore2013,Hinton2007}, and even to study physical systems~\cite{Carleo2017,Nomura2017,Lesieur2017,Weinstein2017,Aoki2016,Huang2017a,Decelle2018,Deng2017,Biamonte2017,Deng2017a}. Besides, connections between RBM and tensor networks have been recently reported.~\cite{Chen2018,Huang2017}  

\indent In this paper, we apply the RBM method to compress the Slater determinants of the FCI ground-state wavefunctions of four singlet molecules: BeH\textsubscript{2}, C\textsubscript{2}, N\textsubscript{2}, and
F\textsubscript{2}. On top of that, we investigate the reduction of the configuration spaces induced by the RBM, and generate potential energy surfaces (PES's) within a chemical accuracy level (1 kcal/mol). By the results, the RBM method sounds to be an alternative approach of lessening the computational cost of the determinant-based CI algorithms.  	

\hypertarget{sec:formalism}{}
\section{Formalism}

\indent Our task is to find a compact representation of the Slater configurations~($\Psi_{n}$'s) that span $\Phi_{\text{FCI}}$ (Eq.~\ref{fci_disc_eq}). Each $\Psi_{n}$'s is binary, since the configurations represent the occupied~(=~1) and virtual~(=~0) spin atomic orbitals~\cite{Szabo1989, Mayer2003,Helgaker2000}. Being the number of spin atomic orbitals predefined by the basis set of the atoms that compose a system.~\cite{Szabo1989,Helgaker2000}

%encoding figure
\begin{figure}[t]
    \hypertarget{fig:encoding_process}{}
    \centering
    \includegraphics[width=0.47\textwidth]{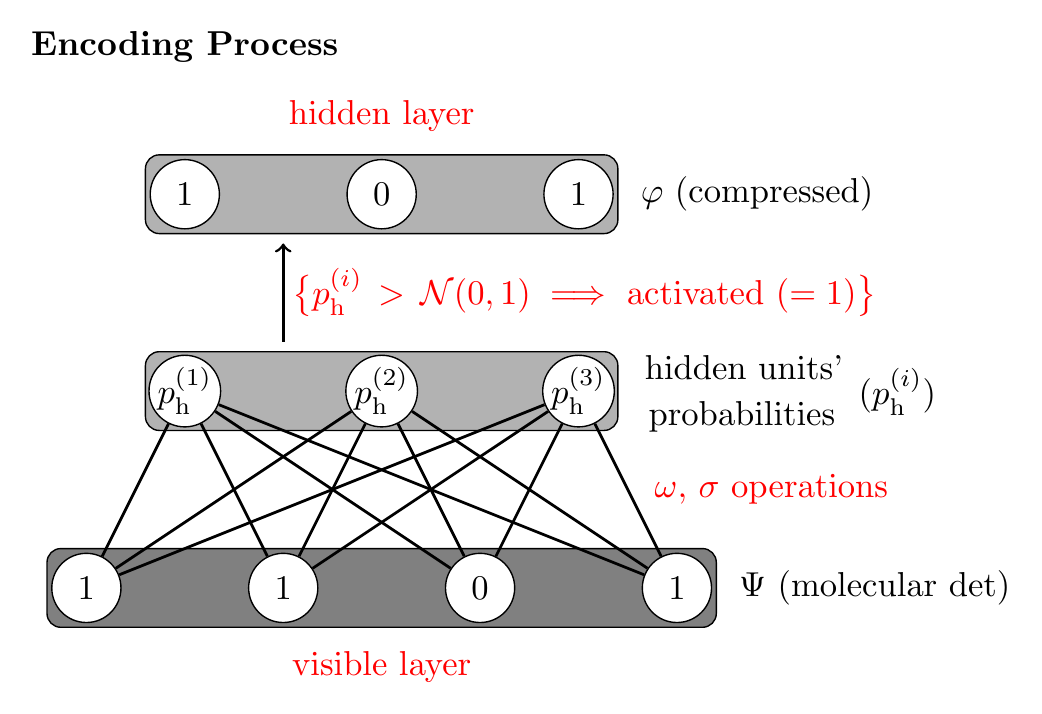} 
    \caption{Encoding Process: a fictitious molecular determinant (det), [1,1,0,1], which has 4 bits, is compressed to [1,0,1], which has 3 bits. $\omega$ is a set of weights that connect the visible and hidden layers, $\sigma$ is a logistic function, $\mathcal{N}(0,1)$ is a normal distribution with mean 0 and variance 1, and ``$\scriptstyle \implies$'' is the ``implies'' symbol. The molecular determinants denote the [occupied~(=1)~/~virtual~(=0)] spin atomic orbitals of a given system. See~\protect\hyperlink{sec:formalism}{\texttt{Formalism}} for details.~\cite{Figures_Paper}}
\end{figure}

\indent Suppose ``$i$'' is a unit of the hidden layer (h) and ``$j$'' is a unit of the visible layer (v). Let $\varphi$ be the compressed configuration associated to  $\Psi$~(a member of the $\Psi_{n}$'s), and $\omega$, a set of weights which connects the visible and the hidden layer. 

\indent The encoding process~(FIG.~\protect\hyperlink{fig:encoding_process}{1}.) can be expressed by $p_{\text{h}}^{(i)}$, the probability of the hidden unit~``$i$'':~\cite{Smolensky1986,Freund1992,Hinton2012,Hinton2006,Bengio2012,Kingma2013,Chen2017,LeRoux2008,Montufar2011}

\begin{equation}
p_{\text{h}}^{(i)}=\sigma\left\{ d_{i}+\sum_{j}\;\left[\Psi\right]_{j}\omega_{ji}\right\}\text{,} 
\end{equation}
where $\sigma\left(t\right)=1/\left[1+\text{exp}\left(-t\right)\right]$~(a logistic function), $d_{i}$~is a bias parameter, and the sum runs over all the ``$j$'' units of $\Psi$.

\indent If $p_{\text{h}}^{(i)}$ is greater than a random number coming from a normal distribution with mean 0 and variance 1, then the hidden unit ``$i$'' is activated (``$i$''~=~1).~\cite{Hinton2012,Chen2017} Otherwise, it is not activated (``$i$''~=~0). As a result of this stochastic process, $\varphi$ is binary like $\Psi$.

\indent Analogously, the decoding process~(FIG.~\protect\hyperlink{fig:decoding_process}{2}.) can be expressed by~$p_{\text{v}}^{(j)}$, the probability of the reconstructed unit~``$j$'':~\cite{Smolensky1986,Freund1992,Hinton2012,Hinton2006,Bengio2012,Kingma2013,Chen2017,LeRoux2008,Montufar2011}

\begin{equation}
p_{\text{v}}^{(j)}=\sigma\left\{ e_{j}+\sum_{i}\;\left[\varphi\right]_{i}\omega_{ij}\right\}\text{,} 
\end{equation}
where $e_{j}$ is a bias parameter and the sum runs over all the ``$i$'' units of $\varphi$.

\indent On the other hand, the activation of the reconstructed visible units goes in another way. To ensure that the reconstructed configurations belong to a given system, the units with the highest $[p_{\text{v}}^{(j)}]$'s become 1---until the total number of electrons of the given system is reached---and then the remaining units become zero.  

\indent From the formalism above, it is important to note that a reconstructed determinant can be generated from more than one different compressed representation. Nevertheless, a compressed representation can recover only one of the original molecular determinants.

%decoding figure
\begin{figure}[t]
    \hypertarget{fig:decoding_process}{}
    \centering
    \includegraphics[width=0.47\textwidth]{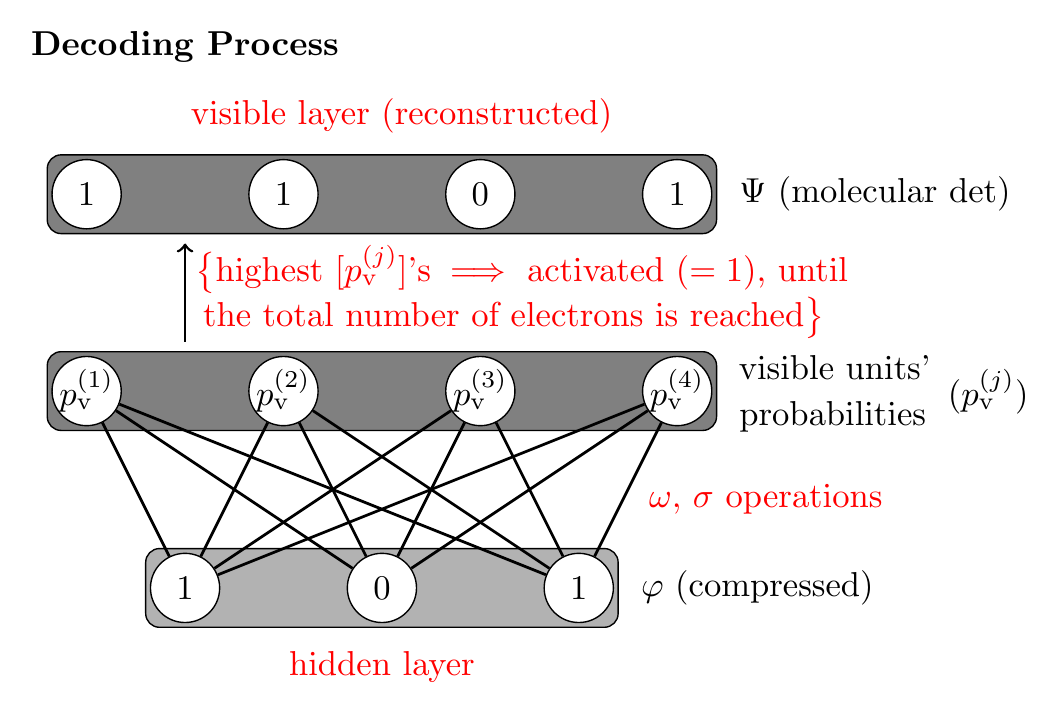} 
    \caption{Decoding Process: starting from the compressed representation, [1,0,1], the same fictitious molecular determinant (det) from Figure ~\protect\hyperlink{fig:encoding_process}{1} is reconstructed. The symbols are defined in Figure~\protect\hyperlink{fig:encoding_process}{1} and in~\protect\hyperlink{sec:formalism}{\texttt{Formalism}}. Observe the distinctiveness between $p_{\text{v}}^{(j)}$ and $p_{\text{h}}^{(i)}$ [in FIG.~\protect\hyperlink{fig:encoding_process}{1}.] to, respectively, reconstruct and compress the molecular det. Because the number of electrons is held constant, the reconstructed determinant will certainly belong to the studied system.~\cite{Figures_Paper}}
\end{figure}

\indent For the next Sections, since the input configurations are molecular determinants, we name this kind of RBM as ``molecular RBM''.

\hypertarget{sec:comp_det}{}
\section{Computational Details} 

\indent STO-3G~\cite{Hehre1969,Feller1996,Schuchardt2007} is the basis set for the four singlet systems studied here: BeH\textsubscript{2}, C\textsubscript{2}, N\textsubscript{2}, and
F\textsubscript{2}. All the electronic structure calculations are performed on PySCF package~\cite{Sun2017}, adopting the L\"{o}wdin-orthogonalized orbitals~\cite{Lowdin1950,Mayer2003}. And, for each system, a molecular RBM is trained by the single-step contrastive divergence algorithm~\cite{Hinton2002, Hinton2012, Chen2017} on a slightly modified version of the Chen \emph{et al.}'s code~\cite{Chen2017}---at the present time, the training is evaluated by the sum of the squared FCI coefficients of not repeated reconstructed configurations, and the units of the reconstructed configurations obey the total number of electrons of a given system to be activated (\emph{vide}~\hyperlink{sec:formalism}{\texttt{Formalism}}).   

\indent Turning to the training set, it follows the alpha and beta string introduced by Handy~\cite{Handy1980,Knowles1984,Sherrill1999,Helgaker2000}, in a manner to guarantee that the determinants are eigenfunctions of~$\hat{S}_{\text{Z}}$ (the z-component of the spin operator)~\cite{Szabo1989,Pauncz1979,Helgaker2000,Roos1994,Shankar1994}. Besides, for each system, the training set is composed of the necessary molecular determinants to recover the ground-state FCI electronic energy---within a chemical accuracy level---of 30 dissociation geometries. These geometries have varying distance~(R), ranging from 0.3 to 3.2 angstrom~(\r{A}), equally spaced by 0.1~\r{A}. 

\indent For all the systems, the dissociation of the molecules into their atoms takes place in one dimension; with a particular attention to the hydrogens in BeH\textsubscript{2}. Both hydrogens are dissociated from the Be atom in an equal fashion. Or, in other words, for each geometry in BeH\textsubscript{2}, the distance~H-Be---which ranges from 0.3 to 3.2~\r{A} in the training set---is identical for the other H atom.

\indent After the molecular RBM is trained, each one of all the $n$th-excited configurations, from the analyzed molecule, is sampled 100 times through the encoding and decoding processes. The decoding process' output with the highest frequency is pointed out as the reconstructed determinant, and the associated encoding output is pointed out as the compressed representation. In the end, the ground-state electronic energy for the molecular RBM is calculated by a projection of the reconstructed determinants onto the FCI determinants, using the Davidson diagonalization method~\cite{Davidson1975,Leininger2001,Sun2017}. 

\indent In this work, the spatial symmetries of the four molecules are not explored.

\hypertarget{sec:assessing_compression}{}
\section{Assessing the Compression}

\indent The amount of bits per molecular determinant is associated to the number of occupied and virtual atomic orbitals for the uncompressed configurations, and to the number of hidden units for the compressed ones.

\indent With this, we consider the following metric to evaluate the compression achieved by the molecular RBM.

\begin{equation}
\text{TNB}=\sum_{s}\;\text{fbits}\left(\varUpsilon_{s}\right)\text{,}
\label{tnb_equation}
\end{equation}
where ``TNB''~=~Total Number of Bits, ``fbits'' is a function which counts the number of bits of the $s$th~compressed/not compressed molecular determinant~$(\varUpsilon_{s})$. And the sum runs through not repeated configurations.

\indent Physically, this metric concatenates all the determinants of a system in the same line and computes the number of bits of this concatenation. Furthermore, the metric above not only consider the compression for each configuration in the CI expansion, but also considers the reduction of the configuration space that span $\Phi_{\text{FCI}}$~(Eq.~\ref{fci_disc_eq}). 

\indent Since the studied systems are singlet, only configurations satisfying~$\langle\hat{S}_{\text{Z}}\rangle=0$ (the expectation value of the $\hat{S}_{\text{Z}}$ operator)~\cite{Szabo1989,Pauncz1979,Helgaker2000,Roos1994,Shankar1994} enter in the metric. Moreover, for the compressed configurations, TNB considers only the minimum compressed representations that recover not repeated uncompressed ones.

\indent Moving to the PES, we consider the nonparallelism error~(NPE)~\cite{Li1995,Paldus2007} to evaluate the potential curve generated by the compression. Within an interval R, NPE is defined by the distance between two points: the greatest and the lowest signed deviations compared to the FCI curve.~\cite{Li1995,Paldus2007} And, for each considered molecule, NPE is calculated for the interval R~$\in$~[0.3, 5.8]~\text{\r{A}}.

\section{Results and Discussion}

\begin{table}[t!]

\caption{Comparing the space savings and the nonparalelism error (NPE) for the molecular RBM~(mRBM) and the spin-adapted~(SA) CCSD(RHF), under the four singlet systems.~\cite{SM2018}}

\noindent \begin{centering}
\begin{tabular}{ccccc}
\hline 
\hline
\noalign{\vskip0.12cm}
\textbf{Systems} & \textbf{BeH\textsubscript{2}} & \textbf{C\textsubscript{2}} & \textbf{N\textsubscript{2}} & \textbf{F\textsubscript{2}}\tabularnewline
\noalign{\vskip0.12cm}
\hline 
\noalign{\vskip0.10cm}
\multicolumn{5}{c}{\textbf{Total number of bits (TNB)~\footnote{(See~\hyperlink{sec:assessing_compression}{\texttt{Assessing the Compression}} for definition.) \label{footnote t2_1}}}}\tabularnewline
\noalign{\vskip0.10cm}
FCI & 17,150 & 882,000 & 288,000 & 2,000\tabularnewline
\noalign{\vskip0.05cm}
SA CCSD(RHF) & 1,274 & 6,500 & 5,060 & 1,100\tabularnewline
\noalign{\vskip0.05cm}
mRBM & 3,260  & 737,064  & 52,845  & 406\tabularnewline
\hline 
\noalign{\vskip0.10cm}
\multicolumn{5}{c}{\textbf{Space savings~\footnote{Space savings $=$ \{1 $-$ TNB(compressed)/TNB(FCI)\}} (\%)}}\tabularnewline
\noalign{\vskip0.10cm}
SA CCSD(RHF) & 92.6 & 99.3 & 98.2 & 45.0\tabularnewline
\noalign{\vskip0.05cm}
mRBM & 81.0 & 16.4 & 81.7 & 79.7\tabularnewline
\hline 
\noalign{\vskip0.10cm}
\multicolumn{5}{c}{\textbf{NPE}~\textsuperscript{\ref{footnote t2_1},}\!~\footnote{For the interval R $\in$ [0.3, 5.8] \text{\r{A}}.} \textbf{(kcal/mol)}}\tabularnewline
\noalign{\vskip0.10cm}
CCSD(RHF) & 4.94 & 38.3  & 144.4 & 0.0\tabularnewline
\noalign{\vskip0.05cm}
mRBM & 0.0 & 0.1 & 0.3 & 0.3\tabularnewline
\hline 
\hline
\noalign{\vskip0.08cm}
\end{tabular}
\par\end{centering}
\label{tab:bits,savings,NPE}
\end{table}

\begin{figure}[t!]
    \hypertarget{fig:BeH2_PES}{}
    \centering
    \includegraphics[width=0.47\textwidth]{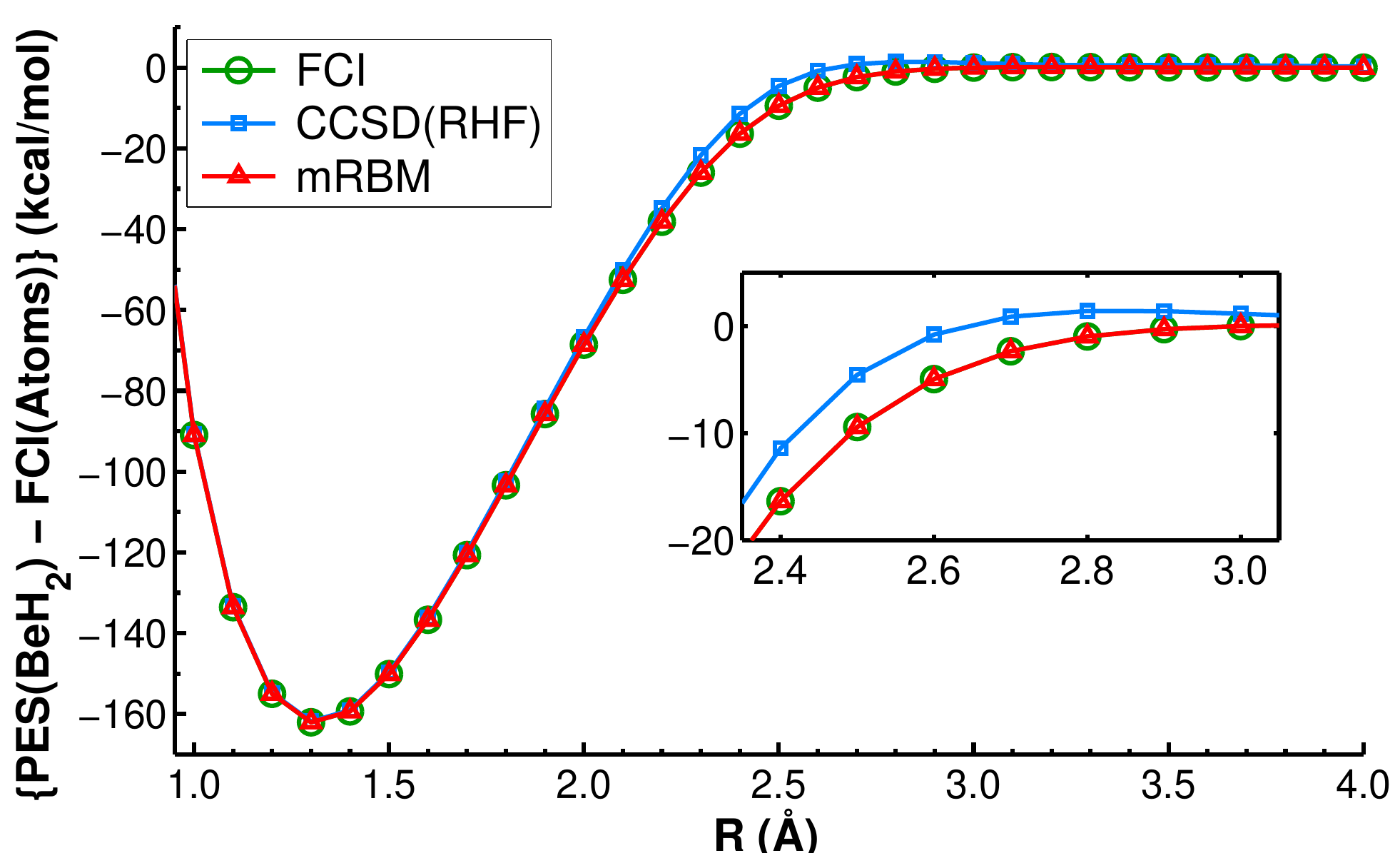} 
    \caption{PES for BeH\textsubscript{2}, subtracting the ground-state FCI electronic energy of one Be and two H atoms from the three curves. (See text for details.)}
\end{figure}

\indent In this section, we abbreviate ``molecular~RBM" to mRBM in tables and graphs. Besides, a comparison with CCSD(RHF) is established. CCSD stands for ``coupled cluster singles~(S) and doubles~(D)", adopting the Restricted Hartree--Fock~(RHF) as the reference determinant for the singly and doubly-excited configurations. The number of bits for CCSD(RHF) is considered under the spin-adapted~(SA) configurations~\cite{Szabo1989,Pauncz1979,Helgaker2000}, which is abbreviated as SA~CCSD(RHF). SA~configurations indicates that each configuration is not only an eigenfunction of~$\hat{S}_{\text{Z}}$---like the uncompressed determinants considered here (see~\hyperlink{sec:comp_det}{\texttt{Computational Details}})---but also an eigenfunction of~$\hat{S}^{2}$ (the total spin-squared operator)~\cite{Szabo1989,Pauncz1979,Helgaker2000,Roos1994,Shankar1994}. 

\indent The unity for energy is kcal/mol and specific aspects of the calculations are in~\hyperlink{sec:comp_det}{\texttt{Computational Details}}. In addition, under the STO-3G basis set, F\textsubscript{2} has only up to doubly-excited configurations, and then CCSD(RHF) becomes exact like FCI. 

\indent Turning to Table~\ref{tab:bits,savings,NPE}, it shows the total number of bits~(TNB), the space savings, and the nonparalelism error~(NPE) for PES---where the distance~(R) between atoms are in the interval [0.3, 5.8]~\text{\r{A}}.

\begin{figure}[b!]
    \hypertarget{fig:C2_PES}{}
    \centering
    \includegraphics[width=0.47\textwidth]{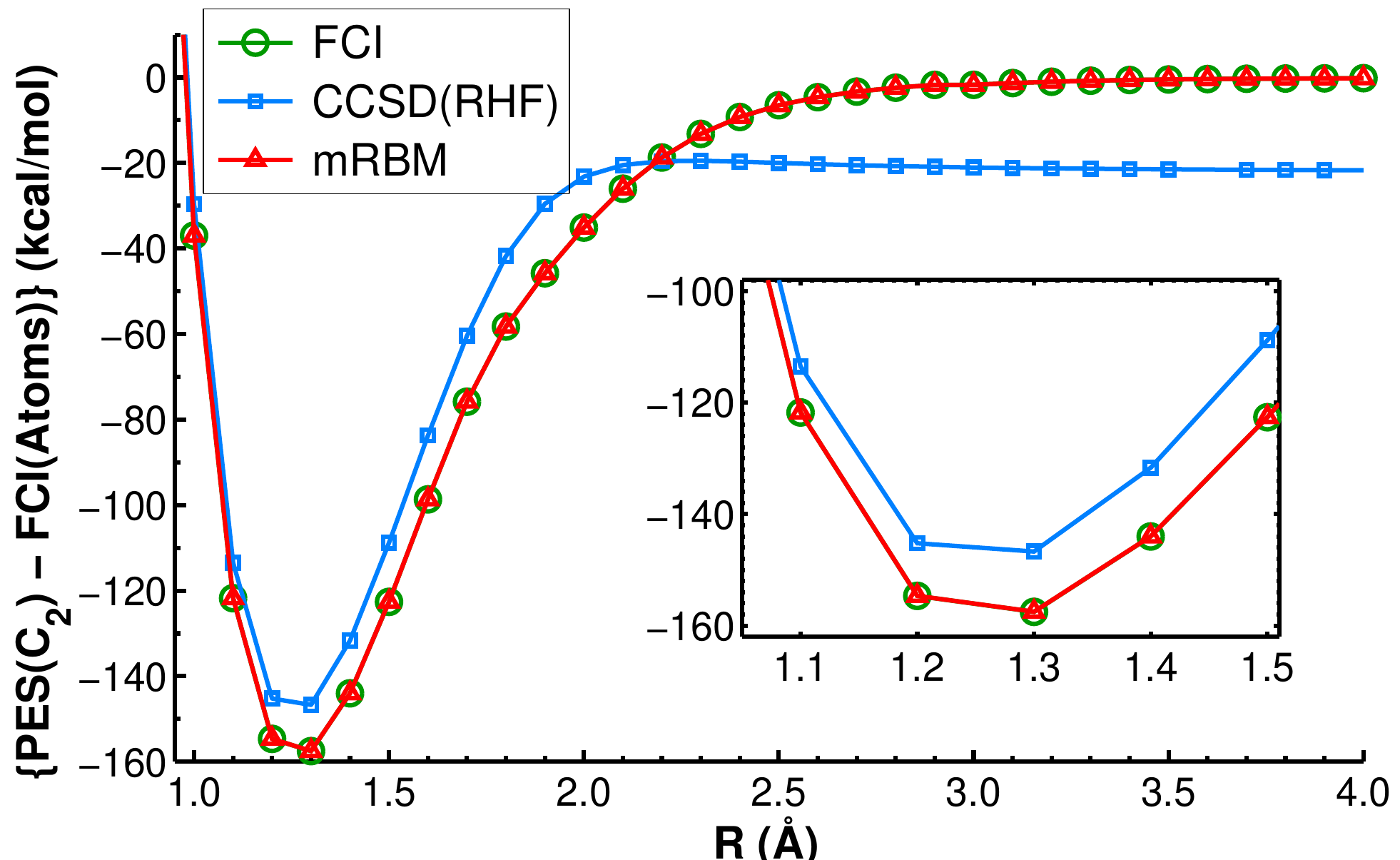}
    \caption{PES for C\textsubscript{2}, subtracting the ground-state FCI electronic energy of two C atoms from the three curves. (See text for details.)}
\end{figure}

\begin{figure}[t]
    \hypertarget{fig:N2_PES}{}
    \centering
    \includegraphics[width=0.47\textwidth]{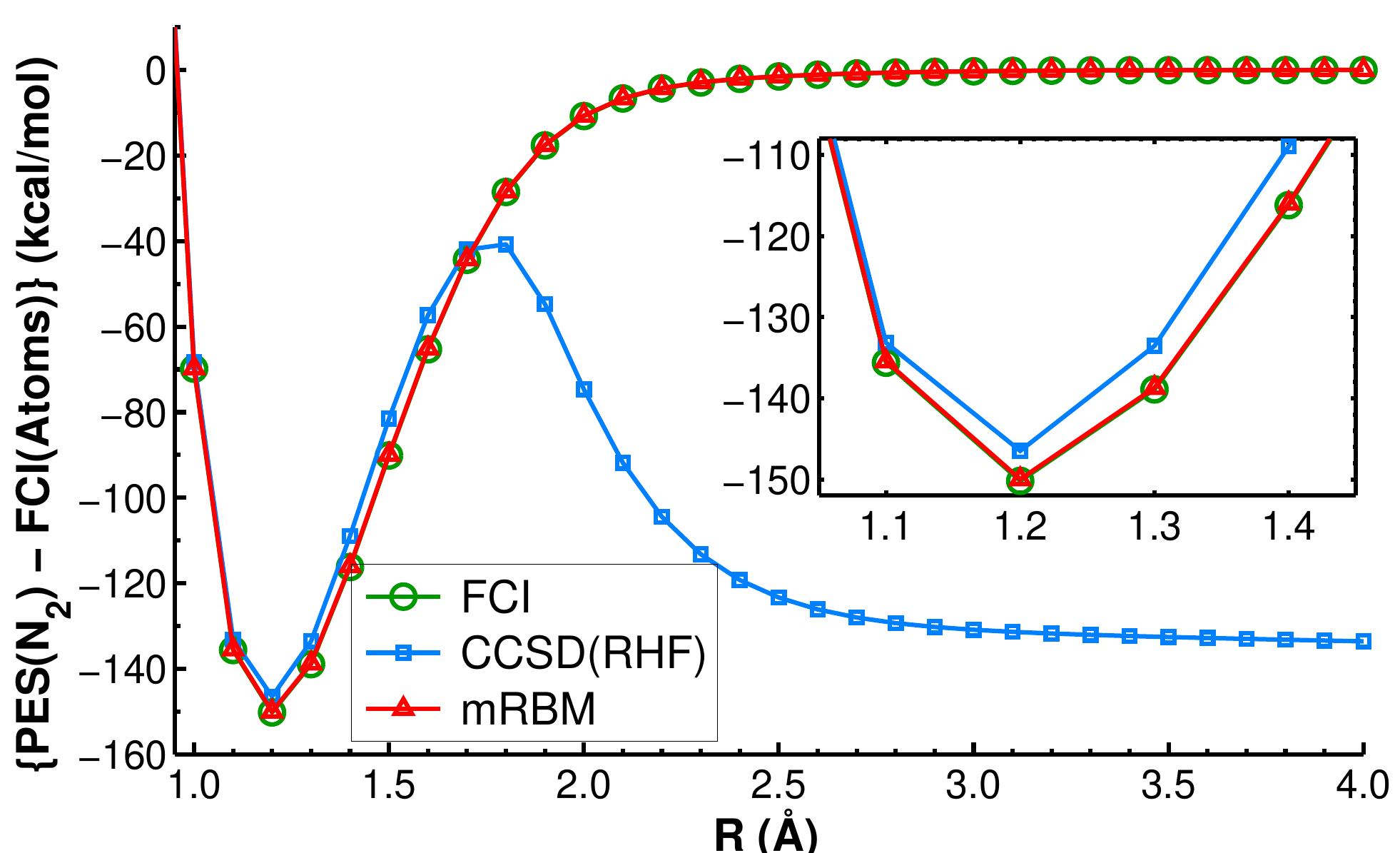}
    \caption{PES for N\textsubscript{2}, subtracting the ground-state FCI electronic energy of two N atoms from the three curves. (See text for details.)}
\end{figure}

\indent TNB is linked to the space savings through the total number of bits of FCI: mRBM and SA~CCSD(RHF) are the compression methods, and are compared to the uncompressed one~(FCI). The space savings for mRBM are in the order of~80\% for BeH\textsubscript{2}, N\textsubscript{2} and F\textsubscript{2}; but it is 16.4\% for C\textsubscript{2}. On the other hand, the space savings for SA~CCSD(RHF) exhibit values of the order of~95\% for BeH\textsubscript{2}, C\textsubscript{2} and N\textsubscript{2}; but it is 45.0\% for F\textsubscript{2}, when CCSD(RHF) is exact, \emph{i.e.}, for SA~CCSD(RHF), the space savings for the singlet F\textsubscript{2} molecule relies just on the SA~configurations embraced.

\indent However, the space savings \emph{per se} does not tell much about how good a compression is, and therefore it must be combined with NPE and PES. Having this in mind, Figure~\hyperlink{fig:BeH2_PES}{{3}} through~\hyperlink{fig:F2_PES}{{6}} display PES's for the four singlet molecules, employing FCI, CCSD(RHF) and mRBM. In each PES, these three curves are subtracted by a constant---the ground-state FCI electronic energy for the atoms that compose a given molecule [PES(molecule)~$-$~FCI(Atoms)].

\begin{figure}[b!]
    \hypertarget{fig:F2_PES}{}
    \centering
    \includegraphics[width=0.47\textwidth]{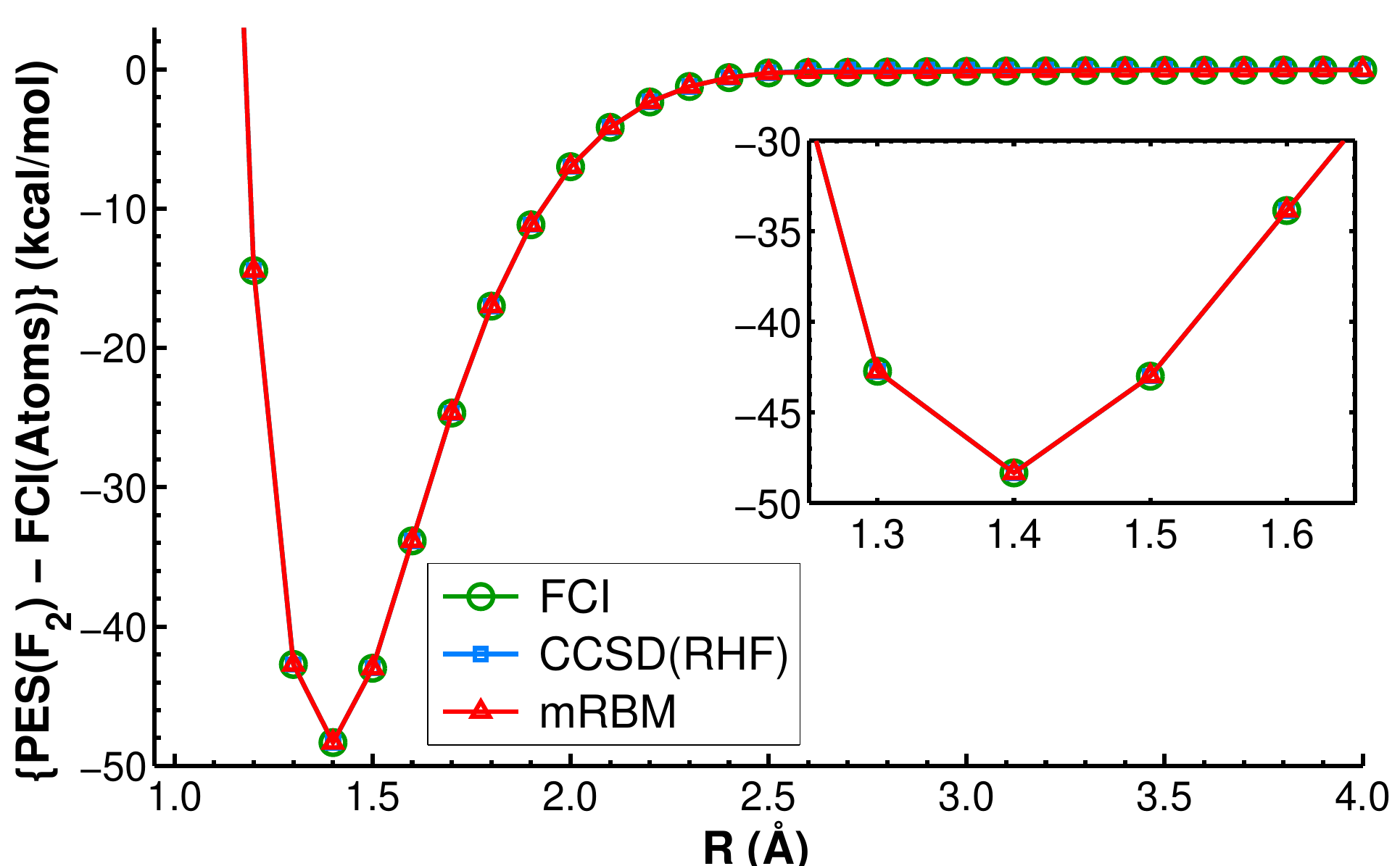}
    \caption{PES for F\textsubscript{2}, subtracting the ground-state FCI electronic energy of two F atoms from the three curves. (See text for details.)}
\end{figure}

\indent For BeH\textsubscript{2}, Figure~\hyperlink{fig:BeH2_PES}{3} shows that CCSD(RHF) diverges from the FCI curve from R~=~1.9 to R~=~3.2~\text{\r{A}}, and it is reflected by BeH\textsubscript{2}'s~NPE value of~4.94~kcal/mol~(Table~\ref{tab:bits,savings,NPE}). The molecular RBM, however, fully recovers the FCI curve, showing a NPE value of~0.0~kcal/mol.

\indent In Figure~\hyperlink{fig:C2_PES}{4}, the CCSD(RHF) curve for C\textsubscript{2} is qualitatively correct until R~=~2.0~\text{\r{A}}. After that point, CCSD(RHF) predicts a lower dissociation energy, characterizing a large NPE value of~38.3~kcal/mol for this system. In its turn, mRBM overlaps FCI, with a lower NPE value of~0.3~kcal/mol. Similarly, Figure~\hyperlink{fig:N2_PES}{5} reveals that the CCSF(RHF) curve for N\textsubscript{2} is qualitatively correct until R~=~1.7~\text{\r{A}}. And, thereafter, it predicts an incorrect dissociation energy. The NPE value for this CCSD(RHF) curve is the largest one in Table~\ref{tab:bits,savings,NPE}:~144.4~kcal/mol. Considering the mRBM, it pratically overlaps the FCI curve, exhibiting a lower NPE value of~0.3~kcal/mol.

\indent The dissociation problem faced by CCSD(RHF), in Figures~\hyperlink{fig:C2_PES}{4} and~\hyperlink{fig:N2_PES}{5}, is known as the size-consistency issue. Due to the RHF reference configuration adopted, this coupled cluster method is not size-consistent in principle.~\cite{Roos1994,Gordon1996,Lee2009}

\indent The last figure---Figure~\hyperlink{fig:F2_PES}{6}---displays the PES for~F\textsubscript{2}. As pointed out before, CCSD(RHF) is exact for this molecule, implying a zero value for NPE. The molecular RBM curve basically lies over FCI as well, but with a NPE value of~0.3~kcal/mol.

\indent In summary, after combining the space savings and NPE from Table~\ref{tab:bits,savings,NPE}, and the four PES's in Figures~\hyperlink{fig:BeH2_PES}{3}--\hyperlink{fig:F2_PES}{6}; the higher compression of CCSD(RHF)---credited for only considering singly and doubly-excited configurations---come at a price: its PES's for BeH\textsubscript{2}, C\textsubscript{2}, and N\textsubscript{2} are not chemical accurate. In contrast, the molecular RBM shows large space savings for BeH\textsubscript{2}, N\textsubscript{2}, and F\textsubscript{2}, and it generates PES's that are chemical accurate for all the four studied molecules.

%\begin{figure*}[t!]
%\centering
 %   \includegraphics[width=0.45\textwidth]{beh2_fci_rbm_pes.eps}\hfil
 %  \includegraphics[width=0.45\textwidth]{c2_fci_rbm_pes.eps}\par\medskip
 %   \includegraphics[width=0.45\textwidth]{n2_fci_rbm_pes.eps}\hfil
 %   \includegraphics[width=0.45\textwidth]{f2_fci_rbm_pes.eps}
 %   \caption{PES - FCI[Atoms] for BeH2, C2, N2, and F2}
%\end{figure*}

\section{Conclusion}

\indent The molecular RBM not only compresses the molecular determinants, but also truncates the FCI~expansion. Because of these facts, mRBM is a possible way of decreasing the computational cost of determinant-driven CI algorithms. Each mRBM includes configurations that are essential for the analyzed system, within a chemical accuracy level, generating smooth~PES and providing space savings that are comparable to the CCSD(RHF) method. 

\indent Different than the coupled cluster, and as a kind of truncated CI expansion, mRBM satisfy the variational theorem~\cite{Szabo1989, Mayer2003,Helgaker2000,Sherrill1999,Shankar1994}, and therefore predicts ground-state energies which are upper bounds of the exact ones. 

\indent Lastly, an atomic version of the RBM---as building blocks for molecules---could increase the compression already achieved by mRBM; and could be a universal approximation to efficiently truncate the FCI~expansion for any~system over any~geometry. These concepts are under investigation and will be compared to the mRBM in the near future. 

\bibliography{mrbm}

\begin{thebibliography}{91}%
\makeatletter
\providecommand \@ifxundefined [1]{%
 \@ifx{#1\undefined}
}%
\providecommand \@ifnum [1]{%
 \ifnum #1\expandafter \@firstoftwo
 \else \expandafter \@secondoftwo
 \fi
}%
\providecommand \@ifx [1]{%
 \ifx #1\expandafter \@firstoftwo
 \else \expandafter \@secondoftwo
 \fi
}%
\providecommand \natexlab [1]{#1}%
\providecommand \enquote  [1]{``#1''}%
\providecommand \bibnamefont  [1]{#1}%
\providecommand \bibfnamefont [1]{#1}%
\providecommand \citenamefont [1]{#1}%
\providecommand \href@noop [0]{\@secondoftwo}%
\providecommand \href [0]{\begingroup \@sanitize@url \@href}%
\providecommand \@href[1]{\@@startlink{#1}\@@href}%
\providecommand \@@href[1]{\endgroup#1\@@endlink}%
\providecommand \@sanitize@url [0]{\catcode `\\12\catcode `\$12\catcode
  `\&12\catcode `\#12\catcode `\^12\catcode `\_12\catcode `\%12\relax}%
\providecommand \@@startlink[1]{}%
\providecommand \@@endlink[0]{}%
\providecommand \url  [0]{\begingroup\@sanitize@url \@url }%
\providecommand \@url [1]{\endgroup\@href {#1}{\urlprefix }}%
\providecommand \urlprefix  [0]{URL }%
\providecommand \Eprint [0]{\href }%
\providecommand \doibase [0]{http://dx.doi.org/}%
\providecommand \selectlanguage [0]{\@gobble}%
\providecommand \bibinfo  [0]{\@secondoftwo}%
\providecommand \bibfield  [0]{\@secondoftwo}%
\providecommand \translation [1]{[#1]}%
\providecommand \BibitemOpen [0]{}%
\providecommand \bibitemStop [0]{}%
\providecommand \bibitemNoStop [0]{.\EOS\space}%
\providecommand \EOS [0]{\spacefactor3000\relax}%
\providecommand \BibitemShut  [1]{\csname bibitem#1\endcsname}%
\let\auto@bib@innerbib\@empty
%</preamble>
\bibitem [{\citenamefont {Kohn}(1999)}]{Kohn1999}%
  \BibitemOpen
  \bibfield  {author} {\bibinfo {author} {\bibfnamefont {W.}~\bibnamefont
  {Kohn}},\ }\href {\doibase 10.1103/RevModPhys.71.1253} {\bibfield  {journal}
  {\bibinfo  {journal} {Rev. Mod. Phys.}\ }\textbf {\bibinfo {volume} {71}},\
  \bibinfo {pages} {1253} (\bibinfo {year} {1999})}\BibitemShut {NoStop}%
\bibitem [{\citenamefont {{Bauschlicher, Charles W.}}\ \emph
  {et~al.}(1989)\citenamefont {{Bauschlicher, Charles W.}}, \citenamefont
  {Langhoff},\ and\ \citenamefont {Taylor}}]{Bauschlicher1989}%
  \BibitemOpen
  \bibfield  {author} {\bibinfo {author} {\bibfnamefont {J.}~\bibnamefont
  {{Bauschlicher, Charles W.}}}, \bibinfo {author} {\bibfnamefont {S.~R.}\
  \bibnamefont {Langhoff}}, \ and\ \bibinfo {author} {\bibfnamefont {P.~R.}\
  \bibnamefont {Taylor}},\ }\href
  {https://ntrs.nasa.gov/search.jsp?R=19900019364} {\emph {\bibinfo {title}
  {{Accurate Quantum Chemical Calculations}}}},\ \bibinfo {type} {Tech. Rep.}\
  \bibinfo {number} {NASA-TM-101932}\ (\bibinfo  {institution} {NASA Ames
  Research Center},\ \bibinfo {year} {1989})\BibitemShut {NoStop}%
\bibitem [{\citenamefont {Roos}(1972)}]{Roos1972}%
  \BibitemOpen
  \bibfield  {author} {\bibinfo {author} {\bibfnamefont {B.}~\bibnamefont
  {Roos}},\ }\href {\doibase 10.1016/0009-2614(72)80140-4} {\bibfield
  {journal} {\bibinfo  {journal} {Chem. Phys. Lett.}\ }\textbf {\bibinfo
  {volume} {15}},\ \bibinfo {pages} {153} (\bibinfo {year} {1972})}\BibitemShut
  {NoStop}%
\bibitem [{\citenamefont {Olsen}\ \emph {et~al.}(1990)\citenamefont {Olsen},
  \citenamefont {J{\o}rgensen},\ and\ \citenamefont {Simons}}]{Olsen1990}%
  \BibitemOpen
  \bibfield  {author} {\bibinfo {author} {\bibfnamefont {J.}~\bibnamefont
  {Olsen}}, \bibinfo {author} {\bibfnamefont {P.}~\bibnamefont {J{\o}rgensen}},
  \ and\ \bibinfo {author} {\bibfnamefont {J.}~\bibnamefont {Simons}},\ }\href
  {\doibase 10.1016/0009-2614(90)85633-N} {\bibfield  {journal} {\bibinfo
  {journal} {Chem. Phys. Lett.}\ }\textbf {\bibinfo {volume} {169}},\ \bibinfo
  {pages} {463} (\bibinfo {year} {1990})}\BibitemShut {NoStop}%
\bibitem [{\citenamefont {Shepard}(1990)}]{Shepard1990}%
  \BibitemOpen
  \bibfield  {author} {\bibinfo {author} {\bibfnamefont {R.}~\bibnamefont
  {Shepard}},\ }\href {\doibase 10.1002/jcc.540110105} {\bibfield  {journal}
  {\bibinfo  {journal} {J. Comput. Chem.}\ }\textbf {\bibinfo {volume} {11}},\
  \bibinfo {pages} {45} (\bibinfo {year} {1990})}\BibitemShut {NoStop}%
\bibitem [{\citenamefont {Szabo}\ and\ \citenamefont
  {Ostlund}(1996)}]{Szabo1989}%
  \BibitemOpen
  \bibfield  {author} {\bibinfo {author} {\bibfnamefont {A.}~\bibnamefont
  {Szabo}}\ and\ \bibinfo {author} {\bibfnamefont {N.~S.}\ \bibnamefont
  {Ostlund}},\ }\href@noop {} {\emph {\bibinfo {title} {Modern Quantum
  Chemistry: Introduction to Advanced Electronic Structure Theory}}},\ Dover
  Books on Chemistry\ (\bibinfo  {publisher} {Dover Publications},\ \bibinfo
  {address} {Mineola, New York},\ \bibinfo {year} {1996})\BibitemShut {NoStop}%
\bibitem [{\citenamefont {Mayer}(2003)}]{Mayer2003}%
  \BibitemOpen
  \bibfield  {author} {\bibinfo {author} {\bibfnamefont {I.}~\bibnamefont
  {Mayer}},\ }\href {\doibase 10.1007/978-1-4757-6519-9} {\emph {\bibinfo
  {title} {{Simple Theorems, Proofs, and Derivations in Quantum Chemistry}}}},\
  Mathematical and Computational Chemistry\ (\bibinfo  {publisher} {Springer
  US},\ \bibinfo {address} {Boston, MA},\ \bibinfo {year} {2003})\BibitemShut
  {NoStop}%
\bibitem [{\citenamefont {Helgaker}\ \emph {et~al.}(2000)\citenamefont
  {Helgaker}, \citenamefont {J{\o}rgensen},\ and\ \citenamefont
  {Olsen}}]{Helgaker2000}%
  \BibitemOpen
  \bibfield  {author} {\bibinfo {author} {\bibfnamefont {T.}~\bibnamefont
  {Helgaker}}, \bibinfo {author} {\bibfnamefont {P.}~\bibnamefont
  {J{\o}rgensen}}, \ and\ \bibinfo {author} {\bibfnamefont {J.}~\bibnamefont
  {Olsen}},\ }\href {\doibase 10.1002/9781119019572} {\emph {\bibinfo {title}
  {{Molecular Electronic-Structure Theory}}}}\ (\bibinfo  {publisher} {John
  Wiley {\&} Sons, Ltd},\ \bibinfo {year} {2000})\BibitemShut {NoStop}%
\bibitem [{\citenamefont {Taylor}(2013)}]{Taylor2013}%
  \BibitemOpen
  \bibfield  {author} {\bibinfo {author} {\bibfnamefont {P.~R.}\ \bibnamefont
  {Taylor}},\ }\href {\doibase 10.1063/1.4816769} {\bibfield  {journal}
  {\bibinfo  {journal} {J. Chem. Phys.}\ }\textbf {\bibinfo {volume} {139}},\
  \bibinfo {pages} {074113} (\bibinfo {year} {2013})}\BibitemShut {NoStop}%
\bibitem [{\citenamefont {McClean}\ and\ \citenamefont
  {Aspuru-Guzik}(2015)}]{McClean2015}%
  \BibitemOpen
  \bibfield  {author} {\bibinfo {author} {\bibfnamefont {J.~R.}\ \bibnamefont
  {McClean}}\ and\ \bibinfo {author} {\bibfnamefont {A.}~\bibnamefont
  {Aspuru-Guzik}},\ }\href {\doibase 10.1039/C5RA23047K} {\bibfield  {journal}
  {\bibinfo  {journal} {RSC Adv.}\ }\textbf {\bibinfo {volume} {5}},\ \bibinfo
  {pages} {102277} (\bibinfo {year} {2015})}\BibitemShut {NoStop}%
\bibitem [{\citenamefont {Knowles}(2015)}]{Knowles2015}%
  \BibitemOpen
  \bibfield  {author} {\bibinfo {author} {\bibfnamefont {P.~J.}\ \bibnamefont
  {Knowles}},\ }\href {\doibase 10.1080/00268976.2014.1003621} {\bibfield
  {journal} {\bibinfo  {journal} {Mol. Phys.}\ }\textbf {\bibinfo {volume}
  {113}},\ \bibinfo {pages} {13} (\bibinfo {year} {2015})}\BibitemShut
  {NoStop}%
\bibitem [{\citenamefont {Marti}\ \emph {et~al.}(2010)\citenamefont {Marti},
  \citenamefont {Bauer}, \citenamefont {Reiher}, \citenamefont {Troyer},\ and\
  \citenamefont {Verstraete}}]{Marti2010}%
  \BibitemOpen
  \bibfield  {author} {\bibinfo {author} {\bibfnamefont {K.~H.}\ \bibnamefont
  {Marti}}, \bibinfo {author} {\bibfnamefont {B.}~\bibnamefont {Bauer}},
  \bibinfo {author} {\bibfnamefont {M.}~\bibnamefont {Reiher}}, \bibinfo
  {author} {\bibfnamefont {M.}~\bibnamefont {Troyer}}, \ and\ \bibinfo {author}
  {\bibfnamefont {F.}~\bibnamefont {Verstraete}},\ }\href {\doibase
  10.1088/1367-2630/12/10/103008} {\bibfield  {journal} {\bibinfo  {journal}
  {New J. Phys.}\ }\textbf {\bibinfo {volume} {12}},\ \bibinfo {pages} {103008}
  (\bibinfo {year} {2010})}\BibitemShut {NoStop}%
\bibitem [{\citenamefont {Buenker}\ and\ \citenamefont
  {Peyerimhoff}(1974)}]{Buenker1974}%
  \BibitemOpen
  \bibfield  {author} {\bibinfo {author} {\bibfnamefont {R.~J.}\ \bibnamefont
  {Buenker}}\ and\ \bibinfo {author} {\bibfnamefont {S.~D.}\ \bibnamefont
  {Peyerimhoff}},\ }\href
  {https://link.springer.com/article/10.1007/PL00020553} {\bibfield  {journal}
  {\bibinfo  {journal} {Theor. Chim. Acta}\ }\textbf {\bibinfo {volume} {35}},\
  \bibinfo {pages} {33} (\bibinfo {year} {1974})}\BibitemShut {NoStop}%
\bibitem [{\citenamefont {Zhang}\ and\ \citenamefont
  {Evangelista}(2016)}]{Zhang2016}%
  \BibitemOpen
  \bibfield  {author} {\bibinfo {author} {\bibfnamefont {T.}~\bibnamefont
  {Zhang}}\ and\ \bibinfo {author} {\bibfnamefont {F.~A.}\ \bibnamefont
  {Evangelista}},\ }\href {\doibase 10.1021/acs.jctc.6b00639} {\bibfield
  {journal} {\bibinfo  {journal} {J. Chem. Theory Comput.}\ }\textbf {\bibinfo
  {volume} {12}},\ \bibinfo {pages} {4326} (\bibinfo {year}
  {2016})}\BibitemShut {NoStop}%
\bibitem [{\citenamefont {Bytautas}\ and\ \citenamefont
  {Ruedenberg}(2004)}]{Bytautas2004}%
  \BibitemOpen
  \bibfield  {author} {\bibinfo {author} {\bibfnamefont {L.}~\bibnamefont
  {Bytautas}}\ and\ \bibinfo {author} {\bibfnamefont {K.}~\bibnamefont
  {Ruedenberg}},\ }\href {https://doi.org/10.1063/1.1811603} {\bibfield
  {journal} {\bibinfo  {journal} {J. Chem. Phys.}\ }\textbf {\bibinfo {volume}
  {121}},\ \bibinfo {pages} {10905} (\bibinfo {year} {2004})}\BibitemShut
  {NoStop}%
\bibitem [{\citenamefont {Bytautas}\ and\ \citenamefont
  {Ruedenberg}(2005)}]{Bytautas2005}%
  \BibitemOpen
  \bibfield  {author} {\bibinfo {author} {\bibfnamefont {L.}~\bibnamefont
  {Bytautas}}\ and\ \bibinfo {author} {\bibfnamefont {K.}~\bibnamefont
  {Ruedenberg}},\ }\href {https://doi.org/10.1063/1.1869493} {\bibfield
  {journal} {\bibinfo  {journal} {J. Chem. Phys.}\ }\textbf {\bibinfo {volume}
  {122}},\ \bibinfo {pages} {154110} (\bibinfo {year} {2005})}\BibitemShut
  {NoStop}%
\bibitem [{\citenamefont {Zimmerman}(2017)}]{Zimmerman2017}%
  \BibitemOpen
  \bibfield  {author} {\bibinfo {author} {\bibfnamefont {P.~M.}\ \bibnamefont
  {Zimmerman}},\ }\href {\doibase 10.1063/1.4977727} {\bibfield  {journal}
  {\bibinfo  {journal} {J. Chem. Phys.}\ }\textbf {\bibinfo {volume} {1461}},\
  \bibinfo {pages} {104102} (\bibinfo {year} {2017})}\BibitemShut {NoStop}%
\bibitem [{\citenamefont {Bytautas}\ \emph {et~al.}(2011)\citenamefont
  {Bytautas}, \citenamefont {Henderson}, \citenamefont {Jim{\'{e}}nez-Hoyos},
  \citenamefont {Ellis},\ and\ \citenamefont {Scuseria}}]{Bytautas2011}%
  \BibitemOpen
  \bibfield  {author} {\bibinfo {author} {\bibfnamefont {L.}~\bibnamefont
  {Bytautas}}, \bibinfo {author} {\bibfnamefont {T.~M.}\ \bibnamefont
  {Henderson}}, \bibinfo {author} {\bibfnamefont {C.~A.}\ \bibnamefont
  {Jim{\'{e}}nez-Hoyos}}, \bibinfo {author} {\bibfnamefont {J.~K.}\
  \bibnamefont {Ellis}}, \ and\ \bibinfo {author} {\bibfnamefont {G.~E.}\
  \bibnamefont {Scuseria}},\ }\href {\doibase 10.1063/1.3613706} {\bibfield
  {journal} {\bibinfo  {journal} {J. Chem. Phys.}\ }\textbf {\bibinfo {volume}
  {135}},\ \bibinfo {pages} {044119} (\bibinfo {year} {2011})}\BibitemShut
  {NoStop}%
\bibitem [{\citenamefont {Alcoba}\ \emph {et~al.}(2014)\citenamefont {Alcoba},
  \citenamefont {Torre}, \citenamefont {Lain}, \citenamefont {O{\~{n}}a},
  \citenamefont {Capuzzi}, \citenamefont {Raemdonck}, \citenamefont
  {Bultinck},\ and\ \citenamefont {Neck}}]{Alcoba2014}%
  \BibitemOpen
  \bibfield  {author} {\bibinfo {author} {\bibfnamefont {D.~R.}\ \bibnamefont
  {Alcoba}}, \bibinfo {author} {\bibfnamefont {A.}~\bibnamefont {Torre}},
  \bibinfo {author} {\bibfnamefont {L.}~\bibnamefont {Lain}}, \bibinfo {author}
  {\bibfnamefont {O.~B.}\ \bibnamefont {O{\~{n}}a}}, \bibinfo {author}
  {\bibfnamefont {P.}~\bibnamefont {Capuzzi}}, \bibinfo {author} {\bibfnamefont
  {M.~V.}\ \bibnamefont {Raemdonck}}, \bibinfo {author} {\bibfnamefont
  {P.}~\bibnamefont {Bultinck}}, \ and\ \bibinfo {author} {\bibfnamefont
  {D.~V.}\ \bibnamefont {Neck}},\ }\href {\doibase 10.1063/1.4904755}
  {\bibfield  {journal} {\bibinfo  {journal} {J. Chem. Phys.}\ }\textbf
  {\bibinfo {volume} {141}},\ \bibinfo {pages} {244118} (\bibinfo {year}
  {2014})}\BibitemShut {NoStop}%
\bibitem [{\citenamefont {Alcoba}\ \emph {et~al.}(2016)\citenamefont {Alcoba},
  \citenamefont {Torre}, \citenamefont {Lain}, \citenamefont {Massaccesi},
  \citenamefont {O{\~{n}}a}, \citenamefont {Ayers}, \citenamefont {Raemdonck},
  \citenamefont {Bultinck},\ and\ \citenamefont {Neck}}]{Alcoba2016}%
  \BibitemOpen
  \bibfield  {author} {\bibinfo {author} {\bibfnamefont {D.~R.}\ \bibnamefont
  {Alcoba}}, \bibinfo {author} {\bibfnamefont {A.}~\bibnamefont {Torre}},
  \bibinfo {author} {\bibfnamefont {L.}~\bibnamefont {Lain}}, \bibinfo {author}
  {\bibfnamefont {G.~E.}\ \bibnamefont {Massaccesi}}, \bibinfo {author}
  {\bibfnamefont {O.~B.}\ \bibnamefont {O{\~{n}}a}}, \bibinfo {author}
  {\bibfnamefont {P.~W.}\ \bibnamefont {Ayers}}, \bibinfo {author}
  {\bibfnamefont {M.~V.}\ \bibnamefont {Raemdonck}}, \bibinfo {author}
  {\bibfnamefont {P.}~\bibnamefont {Bultinck}}, \ and\ \bibinfo {author}
  {\bibfnamefont {D.~V.}\ \bibnamefont {Neck}},\ }\href {\doibase
  10.1007/s00214-016-1905-x} {\bibfield  {journal} {\bibinfo  {journal} {Theor.
  Chem. Acc.}\ }\textbf {\bibinfo {volume} {135}},\ \bibinfo {pages} {153}
  (\bibinfo {year} {2016})}\BibitemShut {NoStop}%
\bibitem [{\citenamefont {Huron}\ \emph {et~al.}(1973)\citenamefont {Huron},
  \citenamefont {Malrieu},\ and\ \citenamefont {Rancurel}}]{Huron1973}%
  \BibitemOpen
  \bibfield  {author} {\bibinfo {author} {\bibfnamefont {B.}~\bibnamefont
  {Huron}}, \bibinfo {author} {\bibfnamefont {J.~P.}\ \bibnamefont {Malrieu}},
  \ and\ \bibinfo {author} {\bibfnamefont {P.}~\bibnamefont {Rancurel}},\
  }\href {\doibase 10.1063/1.1679199} {\bibfield  {journal} {\bibinfo
  {journal} {J. Chem. Phys.}\ }\textbf {\bibinfo {volume} {58}},\ \bibinfo
  {pages} {5745} (\bibinfo {year} {1973})}\BibitemShut {NoStop}%
\bibitem [{\citenamefont {Evangelisti}\ \emph {et~al.}(1983)\citenamefont
  {Evangelisti}, \citenamefont {Daudey},\ and\ \citenamefont
  {Malrieu}}]{Evangelisti1983}%
  \BibitemOpen
  \bibfield  {author} {\bibinfo {author} {\bibfnamefont {S.}~\bibnamefont
  {Evangelisti}}, \bibinfo {author} {\bibfnamefont {J.-P.}\ \bibnamefont
  {Daudey}}, \ and\ \bibinfo {author} {\bibfnamefont {J.-P.}\ \bibnamefont
  {Malrieu}},\ }\href {\doibase 10.1016/0301-0104(83)85011-3} {\bibfield
  {journal} {\bibinfo  {journal} {Chem. Phys.}\ }\textbf {\bibinfo {volume}
  {75}},\ \bibinfo {pages} {91} (\bibinfo {year} {1983})}\BibitemShut {NoStop}%
\bibitem [{\citenamefont {Harrison}(1991)}]{Harrison1991}%
  \BibitemOpen
  \bibfield  {author} {\bibinfo {author} {\bibfnamefont {R.~J.}\ \bibnamefont
  {Harrison}},\ }\href {\doibase 10.1063/1.460537} {\bibfield  {journal}
  {\bibinfo  {journal} {J. Chem. Phys.}\ }\textbf {\bibinfo {volume} {94}},\
  \bibinfo {pages} {5021} (\bibinfo {year} {1991})}\BibitemShut {NoStop}%
\bibitem [{\citenamefont {Roth}(2009)}]{Roth2009}%
  \BibitemOpen
  \bibfield  {author} {\bibinfo {author} {\bibfnamefont {R.}~\bibnamefont
  {Roth}},\ }\href {\doibase 10.1103/PhysRevC.79.064324} {\bibfield  {journal}
  {\bibinfo  {journal} {Phys. Rev. C}\ }\textbf {\bibinfo {volume} {79}},\
  \bibinfo {pages} {064324} (\bibinfo {year} {2009})}\BibitemShut {NoStop}%
\bibitem [{\citenamefont {Holmes}\ \emph
  {et~al.}(2016{\natexlab{a}})\citenamefont {Holmes}, \citenamefont {Tubman},\
  and\ \citenamefont {Umrigar}}]{Holmes2016}%
  \BibitemOpen
  \bibfield  {author} {\bibinfo {author} {\bibfnamefont {A.~A.}\ \bibnamefont
  {Holmes}}, \bibinfo {author} {\bibfnamefont {N.~M.}\ \bibnamefont {Tubman}},
  \ and\ \bibinfo {author} {\bibfnamefont {C.~J.}\ \bibnamefont {Umrigar}},\
  }\href {\doibase 10.1021/acs.jctc.6b00407} {\bibfield  {journal} {\bibinfo
  {journal} {J. Chem. Theory Comput.}\ }\textbf {\bibinfo {volume} {12}},\
  \bibinfo {pages} {3674} (\bibinfo {year} {2016}{\natexlab{a}})}\BibitemShut
  {NoStop}%
\bibitem [{\citenamefont {Tubman}\ \emph {et~al.}(2016)\citenamefont {Tubman},
  \citenamefont {Lee}, \citenamefont {Takeshita}, \citenamefont {Head-Gordon},\
  and\ \citenamefont {Whaley}}]{Tubman2016}%
  \BibitemOpen
  \bibfield  {author} {\bibinfo {author} {\bibfnamefont {N.~M.}\ \bibnamefont
  {Tubman}}, \bibinfo {author} {\bibfnamefont {J.}~\bibnamefont {Lee}},
  \bibinfo {author} {\bibfnamefont {T.~Y.}\ \bibnamefont {Takeshita}}, \bibinfo
  {author} {\bibfnamefont {M.}~\bibnamefont {Head-Gordon}}, \ and\ \bibinfo
  {author} {\bibfnamefont {K.~B.}\ \bibnamefont {Whaley}},\ }\href {\doibase
  10.1063/1.4955109} {\bibfield  {journal} {\bibinfo  {journal} {J. Chem.
  Phys.}\ }\textbf {\bibinfo {volume} {145}},\ \bibinfo {pages} {044112}
  (\bibinfo {year} {2016})}\BibitemShut {NoStop}%
\bibitem [{\citenamefont {Schriber}\ and\ \citenamefont
  {Evangelista}(2016)}]{Schriber2017}%
  \BibitemOpen
  \bibfield  {author} {\bibinfo {author} {\bibfnamefont {J.~B.}\ \bibnamefont
  {Schriber}}\ and\ \bibinfo {author} {\bibfnamefont {F.~A.}\ \bibnamefont
  {Evangelista}},\ }\href {\doibase 10.1063/1.4948308} {\bibfield  {journal}
  {\bibinfo  {journal} {J. Chem. Phys.}\ }\textbf {\bibinfo {volume} {144}},\
  \bibinfo {pages} {161106} (\bibinfo {year} {2016})}\BibitemShut {NoStop}%
\bibitem [{\citenamefont {Booth}\ \emph {et~al.}(2009)\citenamefont {Booth},
  \citenamefont {Thom},\ and\ \citenamefont {Alavi}}]{Booth2009}%
  \BibitemOpen
  \bibfield  {author} {\bibinfo {author} {\bibfnamefont {G.~H.}\ \bibnamefont
  {Booth}}, \bibinfo {author} {\bibfnamefont {A.~J.~W.}\ \bibnamefont {Thom}},
  \ and\ \bibinfo {author} {\bibfnamefont {A.}~\bibnamefont {Alavi}},\ }\href
  {\doibase 10.1063/1.3193710} {\bibfield  {journal} {\bibinfo  {journal} {J.
  Chem. Phys.}\ }\textbf {\bibinfo {volume} {131}},\ \bibinfo {pages} {054106}
  (\bibinfo {year} {2009})}\BibitemShut {NoStop}%
\bibitem [{\citenamefont {Cleland}\ \emph {et~al.}(2010)\citenamefont
  {Cleland}, \citenamefont {Booth},\ and\ \citenamefont {Alavi}}]{Cleland2010}%
  \BibitemOpen
  \bibfield  {author} {\bibinfo {author} {\bibfnamefont {D.}~\bibnamefont
  {Cleland}}, \bibinfo {author} {\bibfnamefont {G.~H.}\ \bibnamefont {Booth}},
  \ and\ \bibinfo {author} {\bibfnamefont {A.}~\bibnamefont {Alavi}},\ }\href
  {\doibase 10.1063/1.3302277} {\bibfield  {journal} {\bibinfo  {journal} {J.
  Chem. Phys.}\ }\textbf {\bibinfo {volume} {132}},\ \bibinfo {pages} {041103}
  (\bibinfo {year} {2010})}\BibitemShut {NoStop}%
\bibitem [{\citenamefont {Booth}\ \emph {et~al.}(2014)\citenamefont {Booth},
  \citenamefont {Smart},\ and\ \citenamefont {Alavi}}]{Booth2014}%
  \BibitemOpen
  \bibfield  {author} {\bibinfo {author} {\bibfnamefont {G.~H.}\ \bibnamefont
  {Booth}}, \bibinfo {author} {\bibfnamefont {S.~D.}\ \bibnamefont {Smart}}, \
  and\ \bibinfo {author} {\bibfnamefont {A.}~\bibnamefont {Alavi}},\ }\href
  {\doibase 10.1080/00268976.2013.877165} {\bibfield  {journal} {\bibinfo
  {journal} {Mol. Phys.}\ }\textbf {\bibinfo {volume} {112}},\ \bibinfo {pages}
  {1855} (\bibinfo {year} {2014})}\BibitemShut {NoStop}%
\bibitem [{\citenamefont {Petruzielo}\ \emph {et~al.}(2012)\citenamefont
  {Petruzielo}, \citenamefont {Holmes}, \citenamefont {Changlani},
  \citenamefont {Nightingale},\ and\ \citenamefont {Umrigar}}]{Petruzielo2012}%
  \BibitemOpen
  \bibfield  {author} {\bibinfo {author} {\bibfnamefont {F.~R.}\ \bibnamefont
  {Petruzielo}}, \bibinfo {author} {\bibfnamefont {A.~A.}\ \bibnamefont
  {Holmes}}, \bibinfo {author} {\bibfnamefont {H.~J.}\ \bibnamefont
  {Changlani}}, \bibinfo {author} {\bibfnamefont {M.~P.}\ \bibnamefont
  {Nightingale}}, \ and\ \bibinfo {author} {\bibfnamefont {C.~J.}\ \bibnamefont
  {Umrigar}},\ }\href {\doibase 10.1103/PhysRevLett.109.230201} {\bibfield
  {journal} {\bibinfo  {journal} {Phys. Rev. Lett.}\ }\textbf {\bibinfo
  {volume} {109}},\ \bibinfo {pages} {230201} (\bibinfo {year}
  {2012})}\BibitemShut {NoStop}%
\bibitem [{\citenamefont {Ten-no}(2013)}]{Ten-no2013}%
  \BibitemOpen
  \bibfield  {author} {\bibinfo {author} {\bibfnamefont {S.}~\bibnamefont
  {Ten-no}},\ }\href {\doibase 10.1063/1.4802766} {\bibfield  {journal}
  {\bibinfo  {journal} {J. Chem. Phys.}\ }\textbf {\bibinfo {volume} {138}},\
  \bibinfo {pages} {164126} (\bibinfo {year} {2013})}\BibitemShut {NoStop}%
\bibitem [{\citenamefont {Greer}(1995)}]{Greer1995}%
  \BibitemOpen
  \bibfield  {author} {\bibinfo {author} {\bibfnamefont {J.~C.}\ \bibnamefont
  {Greer}},\ }\href {https://doi.org/10.1063/1.469756} {\bibfield  {journal}
  {\bibinfo  {journal} {J. Chem. Phys.}\ }\textbf {\bibinfo {volume} {1031}},\
  \bibinfo {pages} {1821} (\bibinfo {year} {1995})}\BibitemShut {NoStop}%
\bibitem [{\citenamefont {Tong}\ \emph {et~al.}(2000)\citenamefont {Tong},
  \citenamefont {Nolan}, \citenamefont {Cheng},\ and\ \citenamefont
  {Greer}}]{Tong2000}%
  \BibitemOpen
  \bibfield  {author} {\bibinfo {author} {\bibfnamefont {L.}~\bibnamefont
  {Tong}}, \bibinfo {author} {\bibfnamefont {M.}~\bibnamefont {Nolan}},
  \bibinfo {author} {\bibfnamefont {T.}~\bibnamefont {Cheng}}, \ and\ \bibinfo
  {author} {\bibfnamefont {J.}~\bibnamefont {Greer}},\ }\href {\doibase
  10.1016/S0010-4655(00)00119-3} {\bibfield  {journal} {\bibinfo  {journal}
  {Comput. Phys. Commun.}\ }\textbf {\bibinfo {volume} {131}},\ \bibinfo
  {pages} {142} (\bibinfo {year} {2000})}\BibitemShut {NoStop}%
\bibitem [{\citenamefont {Coe}\ and\ \citenamefont {Paterson}(2012)}]{Coe2012}%
  \BibitemOpen
  \bibfield  {author} {\bibinfo {author} {\bibfnamefont {J.~P.}\ \bibnamefont
  {Coe}}\ and\ \bibinfo {author} {\bibfnamefont {M.~J.}\ \bibnamefont
  {Paterson}},\ }\href {\doibase 10.1063/1.4767436} {\bibfield  {journal}
  {\bibinfo  {journal} {J. Chem. Phys.}\ }\textbf {\bibinfo {volume} {137}},\
  \bibinfo {pages} {204108} (\bibinfo {year} {2012})}\BibitemShut {NoStop}%
\bibitem [{\citenamefont {Kelly}\ \emph {et~al.}(2014)\citenamefont {Kelly},
  \citenamefont {Perera}, \citenamefont {Bartlett},\ and\ \citenamefont
  {Greer}}]{Kelly2014}%
  \BibitemOpen
  \bibfield  {author} {\bibinfo {author} {\bibfnamefont {T.~P.}\ \bibnamefont
  {Kelly}}, \bibinfo {author} {\bibfnamefont {A.}~\bibnamefont {Perera}},
  \bibinfo {author} {\bibfnamefont {R.~J.}\ \bibnamefont {Bartlett}}, \ and\
  \bibinfo {author} {\bibfnamefont {J.~C.}\ \bibnamefont {Greer}},\ }\href
  {https://doi.org/10.1063/1.4866609} {\bibfield  {journal} {\bibinfo
  {journal} {J. Chem. Phys.}\ }\textbf {\bibinfo {volume} {140}},\ \bibinfo
  {pages} {084114} (\bibinfo {year} {2014})}\BibitemShut {NoStop}%
\bibitem [{\citenamefont {Holmes}\ \emph
  {et~al.}(2016{\natexlab{b}})\citenamefont {Holmes}, \citenamefont
  {Changlani},\ and\ \citenamefont {Umrigar}}]{Holmes2016a}%
  \BibitemOpen
  \bibfield  {author} {\bibinfo {author} {\bibfnamefont {A.~A.}\ \bibnamefont
  {Holmes}}, \bibinfo {author} {\bibfnamefont {H.~J.}\ \bibnamefont
  {Changlani}}, \ and\ \bibinfo {author} {\bibfnamefont {C.~J.}\ \bibnamefont
  {Umrigar}},\ }\href {\doibase 10.1021/acs.jctc.5b01170} {\bibfield  {journal}
  {\bibinfo  {journal} {J. Chem. Theory Comput.}\ }\textbf {\bibinfo {volume}
  {12}},\ \bibinfo {pages} {1561} (\bibinfo {year}
  {2016}{\natexlab{b}})}\BibitemShut {NoStop}%
\bibitem [{\citenamefont {Ohtsuka}\ and\ \citenamefont
  {Hasegawa}(2017)}]{Ohtsuka2017}%
  \BibitemOpen
  \bibfield  {author} {\bibinfo {author} {\bibfnamefont {Y.}~\bibnamefont
  {Ohtsuka}}\ and\ \bibinfo {author} {\bibfnamefont {J.-Y.}\ \bibnamefont
  {Hasegawa}},\ }\href {\doibase 10.1063/1.4993214} {\bibfield  {journal}
  {\bibinfo  {journal} {J. Chem. Phys.}\ }\textbf {\bibinfo {volume} {147}},\
  \bibinfo {pages} {034102} (\bibinfo {year} {2017})}\BibitemShut {NoStop}%
\bibitem [{\citenamefont {Ivanic}\ and\ \citenamefont
  {Ruedenberg}(2001)}]{Ivanic2001}%
  \BibitemOpen
  \bibfield  {author} {\bibinfo {author} {\bibfnamefont {J.}~\bibnamefont
  {Ivanic}}\ and\ \bibinfo {author} {\bibfnamefont {K.}~\bibnamefont
  {Ruedenberg}},\ }\href {\doibase 10.1007/s002140100285} {\bibfield  {journal}
  {\bibinfo  {journal} {Theor. Chem. Acc.}\ }\textbf {\bibinfo {volume}
  {106}},\ \bibinfo {pages} {339} (\bibinfo {year} {2001})}\BibitemShut
  {NoStop}%
\bibitem [{\citenamefont {Bytautas}\ and\ \citenamefont
  {Ruedenberg}(2009)}]{Bytautas2009}%
  \BibitemOpen
  \bibfield  {author} {\bibinfo {author} {\bibfnamefont {L.}~\bibnamefont
  {Bytautas}}\ and\ \bibinfo {author} {\bibfnamefont {K.}~\bibnamefont
  {Ruedenberg}},\ }\href {\doibase 10.1016/J.CHEMPHYS.2008.11.021} {\bibfield
  {journal} {\bibinfo  {journal} {Chem. Phys.}\ }\textbf {\bibinfo {volume}
  {356}},\ \bibinfo {pages} {64} (\bibinfo {year} {2009})}\BibitemShut
  {NoStop}%
\bibitem [{\citenamefont {Azar}\ and\ \citenamefont
  {Head-Gordon}(2015)}]{Azar2015}%
  \BibitemOpen
  \bibfield  {author} {\bibinfo {author} {\bibfnamefont {R.~J.}\ \bibnamefont
  {Azar}}\ and\ \bibinfo {author} {\bibfnamefont {M.}~\bibnamefont
  {Head-Gordon}},\ }\href {\doibase 10.1063/1.4921377} {\bibfield  {journal}
  {\bibinfo  {journal} {J. Chem. Phys.}\ }\textbf {\bibinfo {volume} {142}},\
  \bibinfo {pages} {204101} (\bibinfo {year} {2015})}\BibitemShut {NoStop}%
\bibitem [{\citenamefont {Smolensky}(1986)}]{Smolensky1986}%
  \BibitemOpen
  \bibfield  {author} {\bibinfo {author} {\bibfnamefont {P.}~\bibnamefont
  {Smolensky}},\ }\enquote {\bibinfo {title} {Information processing in
  dynamical systems: Foundations of harmony theory},}\ in\ \href@noop {} {\emph
  {\bibinfo {booktitle} {Parallel Distributed Processing: Explorations in the
  Microstructure of Cognition, Vol. 1: Foundations}}},\ \bibinfo {editor}
  {edited by\ \bibinfo {editor} {\bibfnamefont {D.~E.}\ \bibnamefont
  {Rumelhart}}, \bibinfo {editor} {\bibfnamefont {J.~L.}\ \bibnamefont
  {McClelland}}, \ and\ \bibinfo {editor} {\bibfnamefont {C.}~\bibnamefont {PDP
  Research~Group}}}\ (\bibinfo  {publisher} {MIT Press},\ \bibinfo {address}
  {Cambridge, MA, USA},\ \bibinfo {year} {1986})\ pp.\ \bibinfo {pages}
  {194--281}\BibitemShut {NoStop}%
\bibitem [{\citenamefont {Freund}\ and\ \citenamefont
  {Haussler}(1992)}]{Freund1992}%
  \BibitemOpen
  \bibfield  {author} {\bibinfo {author} {\bibfnamefont {Y.}~\bibnamefont
  {Freund}}\ and\ \bibinfo {author} {\bibfnamefont {D.}~\bibnamefont
  {Haussler}},\ }\enquote {\bibinfo {title} {Unsupervised learning of
  distributions on binary vectors using two layer networks},}\ in\ \href
  {https://papers.nips.cc/paper/535-unsupervised-learning-of-distributions-on-binary-vectors-using-two-layer-networks}
  {\emph {\bibinfo {booktitle} {Advances in Neural Information Processing
  Systems 4}}},\ \bibinfo {editor} {edited by\ \bibinfo {editor} {\bibfnamefont
  {J.~E.}\ \bibnamefont {Moody}}, \bibinfo {editor} {\bibfnamefont {S.~J.}\
  \bibnamefont {Hanson}}, \ and\ \bibinfo {editor} {\bibfnamefont {R.~P.}\
  \bibnamefont {Lippmann}}}\ (\bibinfo  {publisher} {Morgan-Kaufmann},\
  \bibinfo {year} {1992})\ pp.\ \bibinfo {pages} {912--919}\BibitemShut
  {NoStop}%
\bibitem [{\citenamefont {Hinton}(2012)}]{Hinton2012}%
  \BibitemOpen
  \bibfield  {author} {\bibinfo {author} {\bibfnamefont {G.~E.}\ \bibnamefont
  {Hinton}},\ }\enquote {\bibinfo {title} {{A Practical Guide to Training
  Restricted Boltzmann Machines}},}\ in\ \href {\doibase
  10.1007/978-3-642-35289-8_32} {\emph {\bibinfo {booktitle} {Neural Networks:
  Tricks of the Trade: Second Edition}}},\ \bibinfo {editor} {edited by\
  \bibinfo {editor} {\bibfnamefont {G.}~\bibnamefont {Montavon}}, \bibinfo
  {editor} {\bibfnamefont {G.~B.}\ \bibnamefont {Orr}}, \ and\ \bibinfo
  {editor} {\bibfnamefont {K.-R.}\ \bibnamefont {M{\"u}ller}}}\ (\bibinfo
  {publisher} {Springer Berlin Heidelberg},\ \bibinfo {address} {Berlin,
  Heidelberg},\ \bibinfo {year} {2012})\ pp.\ \bibinfo {pages}
  {599--619}\BibitemShut {NoStop}%
\bibitem [{\citenamefont {Hinton}\ \emph {et~al.}(2006)\citenamefont {Hinton},
  \citenamefont {Osindero},\ and\ \citenamefont {Teh}}]{Hinton2006}%
  \BibitemOpen
  \bibfield  {author} {\bibinfo {author} {\bibfnamefont {G.~E.}\ \bibnamefont
  {Hinton}}, \bibinfo {author} {\bibfnamefont {S.}~\bibnamefont {Osindero}}, \
  and\ \bibinfo {author} {\bibfnamefont {Y.-W.}\ \bibnamefont {Teh}},\ }\href
  {\doibase 10.1162/neco.2006.18.7.1527} {\bibfield  {journal} {\bibinfo
  {journal} {Neural Comput.}\ }\textbf {\bibinfo {volume} {18}},\ \bibinfo
  {pages} {1527} (\bibinfo {year} {2006})}\BibitemShut {NoStop}%
\bibitem [{\citenamefont {Chen}\ \emph {et~al.}(2017)\citenamefont {Chen},
  \citenamefont {Chandra}, \citenamefont {Shukela},\ and\ \citenamefont
  {Connors}}]{Chen2017}%
  \BibitemOpen
  \bibfield  {author} {\bibinfo {author} {\bibfnamefont {E.}~\bibnamefont
  {Chen}}, \bibinfo {author} {\bibfnamefont {M.}~\bibnamefont {Chandra}},
  \bibinfo {author} {\bibfnamefont {V.}~\bibnamefont {Shukela}}, \ and\
  \bibinfo {author} {\bibfnamefont {C.}~\bibnamefont {Connors}},\ }\href@noop
  {} {\enquote {\bibinfo {title} {Restricted boltzmann machines in python},}\
  }\bibinfo {howpublished}
  {\url{https://github.com/echen/restricted-boltzmann-machines/blob/master/rbm.py}}
  (\bibinfo {year} {2017})\BibitemShut {NoStop}%
\bibitem [{\citenamefont {{Le Roux}}\ and\ \citenamefont
  {Bengio}(2008)}]{LeRoux2008}%
  \BibitemOpen
  \bibfield  {author} {\bibinfo {author} {\bibfnamefont {N.}~\bibnamefont {{Le
  Roux}}}\ and\ \bibinfo {author} {\bibfnamefont {Y.}~\bibnamefont {Bengio}},\
  }\href {\doibase 10.1162/neco.2008.04-07-510} {\bibfield  {journal} {\bibinfo
   {journal} {Neural Comput.}\ }\textbf {\bibinfo {volume} {20}},\ \bibinfo
  {pages} {1631} (\bibinfo {year} {2008})}\BibitemShut {NoStop}%
\bibitem [{\citenamefont {Montufar}\ and\ \citenamefont
  {Ay}(2011)}]{Montufar2011}%
  \BibitemOpen
  \bibfield  {author} {\bibinfo {author} {\bibfnamefont {G.}~\bibnamefont
  {Montufar}}\ and\ \bibinfo {author} {\bibfnamefont {N.}~\bibnamefont {Ay}},\
  }\href {\doibase 10.1162/NECO\_a\_00113} {\bibfield  {journal} {\bibinfo
  {journal} {Neural Comput.}\ }\textbf {\bibinfo {volume} {23}},\ \bibinfo
  {pages} {1306} (\bibinfo {year} {2011})}\BibitemShut {NoStop}%
\bibitem [{\citenamefont {Carleo}\ and\ \citenamefont
  {Troyer}(2017)}]{Carleo2017}%
  \BibitemOpen
  \bibfield  {author} {\bibinfo {author} {\bibfnamefont {G.}~\bibnamefont
  {Carleo}}\ and\ \bibinfo {author} {\bibfnamefont {M.}~\bibnamefont
  {Troyer}},\ }\href {\doibase 10.1126/science.aag2302} {\bibfield  {journal}
  {\bibinfo  {journal} {Science}\ }\textbf {\bibinfo {volume} {355}},\ \bibinfo
  {pages} {602} (\bibinfo {year} {2017})}\BibitemShut {NoStop}%
\bibitem [{\citenamefont {Bengio}(2012)}]{Bengio2012}%
  \BibitemOpen
  \bibfield  {author} {\bibinfo {author} {\bibfnamefont {Y.}~\bibnamefont
  {Bengio}},\ }\enquote {\bibinfo {title} {{Practical Recommendations for
  Gradient-Based Training of Deep Architectures}},}\ in\ \href {\doibase
  10.1007/978-3-642-35289-8_26} {\emph {\bibinfo {booktitle} {Neural Networks:
  Tricks of the Trade: Second Edition}}},\ \bibinfo {editor} {edited by\
  \bibinfo {editor} {\bibfnamefont {G.}~\bibnamefont {Montavon}}, \bibinfo
  {editor} {\bibfnamefont {G.~B.}\ \bibnamefont {Orr}}, \ and\ \bibinfo
  {editor} {\bibfnamefont {K.-R.}\ \bibnamefont {M{\"u}ller}}}\ (\bibinfo
  {publisher} {Springer Berlin Heidelberg},\ \bibinfo {address} {Berlin,
  Heidelberg},\ \bibinfo {year} {2012})\ pp.\ \bibinfo {pages}
  {437--478}\BibitemShut {NoStop}%
\bibitem [{\citenamefont {Kingma}\ and\ \citenamefont
  {Welling}(2013)}]{Kingma2013}%
  \BibitemOpen
  \bibfield  {author} {\bibinfo {author} {\bibfnamefont {D.~P.}\ \bibnamefont
  {Kingma}}\ and\ \bibinfo {author} {\bibfnamefont {M.}~\bibnamefont
  {Welling}},\ }\href {http://arxiv.org/abs/1312.6114} {\  (\bibinfo {year}
  {2013})},\ \Eprint {http://arxiv.org/abs/1312.6114} {arXiv:1312.6114}
  \BibitemShut {NoStop}%
\bibitem [{\citenamefont {Hinton}\ and\ \citenamefont
  {Salakhutdinov}(2006)}]{Hinton2006a}%
  \BibitemOpen
  \bibfield  {author} {\bibinfo {author} {\bibfnamefont {G.~E.}\ \bibnamefont
  {Hinton}}\ and\ \bibinfo {author} {\bibfnamefont {R.~R.}\ \bibnamefont
  {Salakhutdinov}},\ }\href {\doibase 10.1126/science.1127647} {\bibfield
  {journal} {\bibinfo  {journal} {Science}\ }\textbf {\bibinfo {volume}
  {313}},\ \bibinfo {pages} {504} (\bibinfo {year} {2006})}\BibitemShut
  {NoStop}%
\bibitem [{\citenamefont {Torralba}\ \emph {et~al.}(2008)\citenamefont
  {Torralba}, \citenamefont {Fergus},\ and\ \citenamefont
  {Weiss}}]{Torralba2008}%
  \BibitemOpen
  \bibfield  {author} {\bibinfo {author} {\bibfnamefont {A.}~\bibnamefont
  {Torralba}}, \bibinfo {author} {\bibfnamefont {R.}~\bibnamefont {Fergus}}, \
  and\ \bibinfo {author} {\bibfnamefont {Y.}~\bibnamefont {Weiss}},\ }\enquote
  {\bibinfo {title} {{Small Codes and Large Image Databases for
  Recognition}},}\ in\ \href {\doibase 10.1109/CVPR.2008.4587633} {\emph
  {\bibinfo {booktitle} {2008 IEEE Conf. Comput. Vis. Pattern Recognit.}}}\
  (\bibinfo  {publisher} {IEEE},\ \bibinfo {year} {2008})\ pp.\ \bibinfo
  {pages} {1--8}\BibitemShut {NoStop}%
\bibitem [{\citenamefont {Cheng}\ \emph {et~al.}(2016)\citenamefont {Cheng},
  \citenamefont {Wang}, \citenamefont {Chen},\ and\ \citenamefont
  {Yang}}]{Cheng2016}%
  \BibitemOpen
  \bibfield  {author} {\bibinfo {author} {\bibfnamefont {C.}~\bibnamefont
  {Cheng}}, \bibinfo {author} {\bibfnamefont {S.}~\bibnamefont {Wang}},
  \bibinfo {author} {\bibfnamefont {X.}~\bibnamefont {Chen}}, \ and\ \bibinfo
  {author} {\bibfnamefont {Y.}~\bibnamefont {Yang}},\ }\href {\doibase
  10.1155/2016/1851829} {\bibfield  {journal} {\bibinfo  {journal} {Int. J.
  Distrib. Sens. Networks}\ }\textbf {\bibinfo {volume} {12}},\ \bibinfo
  {pages} {1851829} (\bibinfo {year} {2016})}\BibitemShut {NoStop}%
\bibitem [{\citenamefont {Lokare}\ \emph {et~al.}(2015)\citenamefont {Lokare},
  \citenamefont {Birari},\ and\ \citenamefont {Patil}}]{Lokare2015}%
  \BibitemOpen
  \bibfield  {author} {\bibinfo {author} {\bibfnamefont {V.}~\bibnamefont
  {Lokare}}, \bibinfo {author} {\bibfnamefont {S.}~\bibnamefont {Birari}}, \
  and\ \bibinfo {author} {\bibfnamefont {O.}~\bibnamefont {Patil}},\ }\href
  {http://ijcsit.com/docs/Volume 6/vol6issue05/ijcsit20150605139.pdf}
  {\bibfield  {journal} {\bibinfo  {journal} {Int. J. Comput. Sci. Inf.
  Technol.}\ }\textbf {\bibinfo {volume} {6}},\ \bibinfo {pages} {4799}
  (\bibinfo {year} {2015})}\BibitemShut {NoStop}%
\bibitem [{\citenamefont {Salakhutdinov}\ \emph {et~al.}(2007)\citenamefont
  {Salakhutdinov}, \citenamefont {Mnih},\ and\ \citenamefont
  {Hinton}}]{Salakhutdinov2007}%
  \BibitemOpen
  \bibfield  {author} {\bibinfo {author} {\bibfnamefont {R.}~\bibnamefont
  {Salakhutdinov}}, \bibinfo {author} {\bibfnamefont {A.}~\bibnamefont {Mnih}},
  \ and\ \bibinfo {author} {\bibfnamefont {G.}~\bibnamefont {Hinton}},\
  }\enquote {\bibinfo {title} {{Restricted Boltzmann Machines for Collaborative
  Filtering}},}\ in\ \href {\doibase 10.1145/1273496.1273596} {\emph {\bibinfo
  {booktitle} {Proc. 24th Int. Conf. Mach. Learn. - ICML '07}}}\ (\bibinfo
  {publisher} {ACM Press},\ \bibinfo {address} {New York, New York, USA},\
  \bibinfo {year} {2007})\ pp.\ \bibinfo {pages} {791--798}\BibitemShut
  {NoStop}%
\bibitem [{\citenamefont {Larochelle}\ and\ \citenamefont
  {Bengio}(2008)}]{Larochelle2008}%
  \BibitemOpen
  \bibfield  {author} {\bibinfo {author} {\bibfnamefont {H.}~\bibnamefont
  {Larochelle}}\ and\ \bibinfo {author} {\bibfnamefont {Y.}~\bibnamefont
  {Bengio}},\ }\enquote {\bibinfo {title} {{Classification using Discriminative
  Restricted Boltzmann Machines}},}\ in\ \href {\doibase
  10.1145/1390156.1390224} {\emph {\bibinfo {booktitle} {Proc. 25th Int. Conf.
  Mach. Learn. - ICML '08}}}\ (\bibinfo  {publisher} {ACM Press},\ \bibinfo
  {address} {New York, New York, USA},\ \bibinfo {year} {2008})\ pp.\ \bibinfo
  {pages} {536--543}\BibitemShut {NoStop}%
\bibitem [{\citenamefont {Xing}\ \emph {et~al.}(2012)\citenamefont {Xing},
  \citenamefont {Yan},\ and\ \citenamefont {Hauptmann}}]{Xing2012}%
  \BibitemOpen
  \bibfield  {author} {\bibinfo {author} {\bibfnamefont {E.~P.}\ \bibnamefont
  {Xing}}, \bibinfo {author} {\bibfnamefont {R.}~\bibnamefont {Yan}}, \ and\
  \bibinfo {author} {\bibfnamefont {A.~G.}\ \bibnamefont {Hauptmann}},\ }\href
  {http://arxiv.org/abs/1207.1423} {\  (\bibinfo {year} {2012})},\ \bibinfo
  {note} {appears in Proceedings of the Twenty-First Conference on Uncertainty
  in Artificial Intelligence (UAI2005)},\ \Eprint
  {http://arxiv.org/abs/1207.1423} {arXiv:1207.1423} \BibitemShut {NoStop}%
\bibitem [{\citenamefont {Fiore}\ \emph {et~al.}(2013)\citenamefont {Fiore},
  \citenamefont {Palmieri}, \citenamefont {Castiglione},\ and\ \citenamefont
  {{De Santis}}}]{Fiore2013}%
  \BibitemOpen
  \bibfield  {author} {\bibinfo {author} {\bibfnamefont {U.}~\bibnamefont
  {Fiore}}, \bibinfo {author} {\bibfnamefont {F.}~\bibnamefont {Palmieri}},
  \bibinfo {author} {\bibfnamefont {A.}~\bibnamefont {Castiglione}}, \ and\
  \bibinfo {author} {\bibfnamefont {A.}~\bibnamefont {{De Santis}}},\ }\href
  {\doibase 10.1016/J.NEUCOM.2012.11.050} {\bibfield  {journal} {\bibinfo
  {journal} {Neurocomputing}\ }\textbf {\bibinfo {volume} {122}},\ \bibinfo
  {pages} {13} (\bibinfo {year} {2013})}\BibitemShut {NoStop}%
\bibitem [{\citenamefont {Salakhutdinov}\ and\ \citenamefont
  {Hinton}(2007)}]{Hinton2007}%
  \BibitemOpen
  \bibfield  {author} {\bibinfo {author} {\bibfnamefont {R.}~\bibnamefont
  {Salakhutdinov}}\ and\ \bibinfo {author} {\bibfnamefont {G.}~\bibnamefont
  {Hinton}},\ }\enquote {\bibinfo {title} {{Learning a Nonlinear Embedding by
  Preserving Class Neighbourhood Structure}},}\ in\ \href
  {http://proceedings.mlr.press/v2/salakhutdinov07a.html} {\emph {\bibinfo
  {booktitle} {Proceedings of the Eleventh International Conference on
  Artificial Intelligence and Statistics}}},\ \bibinfo {series} {Proceedings of
  Machine Learning Research}, Vol.~\bibinfo {volume} {2},\ \bibinfo {editor}
  {edited by\ \bibinfo {editor} {\bibfnamefont {M.}~\bibnamefont {Meila}}\ and\
  \bibinfo {editor} {\bibfnamefont {X.}~\bibnamefont {Shen}}}\ (\bibinfo
  {publisher} {PMLR},\ \bibinfo {address} {San Juan, Puerto Rico},\ \bibinfo
  {year} {2007})\ pp.\ \bibinfo {pages} {412--419}\BibitemShut {NoStop}%
\bibitem [{\citenamefont {Nomura}\ \emph {et~al.}(2017)\citenamefont {Nomura},
  \citenamefont {Darmawan}, \citenamefont {Yamaji},\ and\ \citenamefont
  {Imada}}]{Nomura2017}%
  \BibitemOpen
  \bibfield  {author} {\bibinfo {author} {\bibfnamefont {Y.}~\bibnamefont
  {Nomura}}, \bibinfo {author} {\bibfnamefont {A.~S.}\ \bibnamefont
  {Darmawan}}, \bibinfo {author} {\bibfnamefont {Y.}~\bibnamefont {Yamaji}}, \
  and\ \bibinfo {author} {\bibfnamefont {M.}~\bibnamefont {Imada}},\ }\href
  {https://doi.org/10.1103/PhysRevB.96.205152} {\bibfield  {journal} {\bibinfo
  {journal} {Phys. Rev. B}\ }\textbf {\bibinfo {volume} {96}} (\bibinfo {year}
  {2017})}\BibitemShut {NoStop}%
\bibitem [{\citenamefont {Lesieur}\ \emph {et~al.}(2017)\citenamefont
  {Lesieur}, \citenamefont {Krzakala}, \citenamefont {Zdeborov{\'{a}}},
  \citenamefont {Benedetti}, \citenamefont {Realpe-G{\'{o}}mez}, \citenamefont
  {Perdomo-Ortiz}, \citenamefont {Decelle}, \citenamefont {Fissore},\ and\
  \citenamefont {Furtlehner}}]{Lesieur2017}%
  \BibitemOpen
  \bibfield  {author} {\bibinfo {author} {\bibfnamefont {T.}~\bibnamefont
  {Lesieur}}, \bibinfo {author} {\bibfnamefont {F.}~\bibnamefont {Krzakala}},
  \bibinfo {author} {\bibfnamefont {L.}~\bibnamefont {Zdeborov{\'{a}}}},
  \bibinfo {author} {\bibfnamefont {M.}~\bibnamefont {Benedetti}}, \bibinfo
  {author} {\bibfnamefont {J.}~\bibnamefont {Realpe-G{\'{o}}mez}}, \bibinfo
  {author} {\bibfnamefont {A.}~\bibnamefont {Perdomo-Ortiz}}, \bibinfo {author}
  {\bibfnamefont {A.}~\bibnamefont {Decelle}}, \bibinfo {author} {\bibfnamefont
  {G.}~\bibnamefont {Fissore}}, \ and\ \bibinfo {author} {\bibfnamefont
  {C.}~\bibnamefont {Furtlehner}},\ }\href
  {http://iopscience.iop.org/article/10.1209/0295-5075/119/60001/meta}
  {\bibfield  {journal} {\bibinfo  {journal} {EPL}\ }\textbf {\bibinfo {volume}
  {119}} (\bibinfo {year} {2017})}\BibitemShut {NoStop}%
\bibitem [{\citenamefont {Weinstein}(2017)}]{Weinstein2017}%
  \BibitemOpen
  \bibfield  {author} {\bibinfo {author} {\bibfnamefont {S.}~\bibnamefont
  {Weinstein}},\ }\href {https://arxiv.org/abs/1707.03114} {\  (\bibinfo {year}
  {2017})},\ \Eprint {http://arxiv.org/abs/1707.03114} {arXiv:1707.03114}
  \BibitemShut {NoStop}%
\bibitem [{\citenamefont {Aoki}\ and\ \citenamefont
  {Kobayashi}(2016)}]{Aoki2016}%
  \BibitemOpen
  \bibfield  {author} {\bibinfo {author} {\bibfnamefont {K.-I.}\ \bibnamefont
  {Aoki}}\ and\ \bibinfo {author} {\bibfnamefont {T.}~\bibnamefont
  {Kobayashi}},\ }\href {\doibase 10.1142/S0217984916504017} {\bibfield
  {journal} {\bibinfo  {journal} {Mod. Phys. Lett. B}\ }\textbf {\bibinfo
  {volume} {30}},\ \bibinfo {pages} {1650401} (\bibinfo {year}
  {2016})}\BibitemShut {NoStop}%
\bibitem [{\citenamefont {Huang}\ and\ \citenamefont
  {Wang}(2017)}]{Huang2017a}%
  \BibitemOpen
  \bibfield  {author} {\bibinfo {author} {\bibfnamefont {L.}~\bibnamefont
  {Huang}}\ and\ \bibinfo {author} {\bibfnamefont {L.}~\bibnamefont {Wang}},\
  }\href {http://dx.doi.org/10.1103/PhysRevB.95.035105} {\bibfield  {journal}
  {\bibinfo  {journal} {Phys. Rev. B}\ }\textbf {\bibinfo {volume} {95}}
  (\bibinfo {year} {2017})}\BibitemShut {NoStop}%
\bibitem [{\citenamefont {Decelle}\ \emph {et~al.}(2018)\citenamefont
  {Decelle}, \citenamefont {Fissore},\ and\ \citenamefont
  {Furtlehner}}]{Decelle2018}%
  \BibitemOpen
  \bibfield  {author} {\bibinfo {author} {\bibfnamefont {A.}~\bibnamefont
  {Decelle}}, \bibinfo {author} {\bibfnamefont {G.}~\bibnamefont {Fissore}}, \
  and\ \bibinfo {author} {\bibfnamefont {C.}~\bibnamefont {Furtlehner}},\
  }\href {https://arxiv.org/abs/1803.01960v1} {\  (\bibinfo {year} {2018})},\
  \Eprint {http://arxiv.org/abs/1803.01960} {arXiv:1803.01960} \BibitemShut
  {NoStop}%
\bibitem [{\citenamefont {Deng}\ \emph
  {et~al.}(2017{\natexlab{a}})\citenamefont {Deng}, \citenamefont {Li},\ and\
  \citenamefont {{Das Sarma}}}]{Deng2017}%
  \BibitemOpen
  \bibfield  {author} {\bibinfo {author} {\bibfnamefont {D.-L.}\ \bibnamefont
  {Deng}}, \bibinfo {author} {\bibfnamefont {X.}~\bibnamefont {Li}}, \ and\
  \bibinfo {author} {\bibfnamefont {S.}~\bibnamefont {{Das Sarma}}},\ }\href
  {\doibase 10.1103/PhysRevB.96.195145} {\bibfield  {journal} {\bibinfo
  {journal} {Phys. Rev. B}\ }\textbf {\bibinfo {volume} {96}},\ \bibinfo
  {pages} {195145} (\bibinfo {year} {2017}{\natexlab{a}})}\BibitemShut
  {NoStop}%
\bibitem [{\citenamefont {Biamonte}\ \emph {et~al.}(2017)\citenamefont
  {Biamonte}, \citenamefont {Wittek}, \citenamefont {Pancotti}, \citenamefont
  {Rebentrost}, \citenamefont {Wiebe},\ and\ \citenamefont
  {Lloyd}}]{Biamonte2017}%
  \BibitemOpen
  \bibfield  {author} {\bibinfo {author} {\bibfnamefont {J.}~\bibnamefont
  {Biamonte}}, \bibinfo {author} {\bibfnamefont {P.}~\bibnamefont {Wittek}},
  \bibinfo {author} {\bibfnamefont {N.}~\bibnamefont {Pancotti}}, \bibinfo
  {author} {\bibfnamefont {P.}~\bibnamefont {Rebentrost}}, \bibinfo {author}
  {\bibfnamefont {N.}~\bibnamefont {Wiebe}}, \ and\ \bibinfo {author}
  {\bibfnamefont {S.}~\bibnamefont {Lloyd}},\ }\href {\doibase
  10.1038/nature23474} {\bibfield  {journal} {\bibinfo  {journal} {Nature}\
  }\textbf {\bibinfo {volume} {549}},\ \bibinfo {pages} {195} (\bibinfo {year}
  {2017})}\BibitemShut {NoStop}%
\bibitem [{\citenamefont {Deng}\ \emph
  {et~al.}(2017{\natexlab{b}})\citenamefont {Deng}, \citenamefont {Li},\ and\
  \citenamefont {Das~Sarma}}]{Deng2017a}%
  \BibitemOpen
  \bibfield  {author} {\bibinfo {author} {\bibfnamefont {D.-L.}\ \bibnamefont
  {Deng}}, \bibinfo {author} {\bibfnamefont {X.}~\bibnamefont {Li}}, \ and\
  \bibinfo {author} {\bibfnamefont {S.}~\bibnamefont {Das~Sarma}},\ }\href
  {\doibase 10.1103/PhysRevX.7.021021} {\bibfield  {journal} {\bibinfo
  {journal} {Phys. Rev. X}\ }\textbf {\bibinfo {volume} {7}},\ \bibinfo {pages}
  {021021} (\bibinfo {year} {2017}{\natexlab{b}})}\BibitemShut {NoStop}%
\bibitem [{\citenamefont {Chen}\ \emph {et~al.}(2018)\citenamefont {Chen},
  \citenamefont {Cheng}, \citenamefont {Xie}, \citenamefont {Wang},\ and\
  \citenamefont {Xiang}}]{Chen2018}%
  \BibitemOpen
  \bibfield  {author} {\bibinfo {author} {\bibfnamefont {J.}~\bibnamefont
  {Chen}}, \bibinfo {author} {\bibfnamefont {S.}~\bibnamefont {Cheng}},
  \bibinfo {author} {\bibfnamefont {H.}~\bibnamefont {Xie}}, \bibinfo {author}
  {\bibfnamefont {L.}~\bibnamefont {Wang}}, \ and\ \bibinfo {author}
  {\bibfnamefont {T.}~\bibnamefont {Xiang}},\ }\href
  {https://doi.org/10.1103/PhysRevB.97.085104} {\bibfield  {journal} {\bibinfo
  {journal} {Phys. Rev. B}\ }\textbf {\bibinfo {volume} {97}} (\bibinfo {year}
  {2018})}\BibitemShut {NoStop}%
\bibitem [{\citenamefont {Huang}\ and\ \citenamefont
  {Moore}(2017)}]{Huang2017}%
  \BibitemOpen
  \bibfield  {author} {\bibinfo {author} {\bibfnamefont {Y.}~\bibnamefont
  {Huang}}\ and\ \bibinfo {author} {\bibfnamefont {J.~E.}\ \bibnamefont
  {Moore}},\ }\href {http://arxiv.org/abs/1701.06246} {\  (\bibinfo {year}
  {2017})},\ \Eprint {http://arxiv.org/abs/1701.06246} {arXiv:1701.06246}
  \BibitemShut {NoStop}%
\bibitem [{Fig()}]{Figures_Paper}%
  \BibitemOpen
  \href@noop {} {}\bibinfo {note} {This figure was constructed using
  Ti\emph{k}Z, from a modified example at
  \url{https://github.com/MartinThoma/LaTeX-examples/tree/master/tikz}}\BibitemShut
  {NoStop}%
\bibitem [{\citenamefont {Hehre}\ \emph {et~al.}(1969)\citenamefont {Hehre},
  \citenamefont {Stewart},\ and\ \citenamefont {Pople}}]{Hehre1969}%
  \BibitemOpen
  \bibfield  {author} {\bibinfo {author} {\bibfnamefont {W.~J.}\ \bibnamefont
  {Hehre}}, \bibinfo {author} {\bibfnamefont {R.~F.}\ \bibnamefont {Stewart}},
  \ and\ \bibinfo {author} {\bibfnamefont {J.~A.}\ \bibnamefont {Pople}},\
  }\href {\doibase 10.1063/1.1672392} {\bibfield  {journal} {\bibinfo
  {journal} {J. Chem. Phys.}\ }\textbf {\bibinfo {volume} {51}},\ \bibinfo
  {pages} {2657} (\bibinfo {year} {1969})}\BibitemShut {NoStop}%
\bibitem [{\citenamefont {Feller}(1996)}]{Feller1996}%
  \BibitemOpen
  \bibfield  {author} {\bibinfo {author} {\bibfnamefont {D.}~\bibnamefont
  {Feller}},\ }\href {\doibase
  10.1002/(SICI)1096-987X(199610)17:13<1571::AID-JCC9>3.0.CO;2-P} {\bibfield
  {journal} {\bibinfo  {journal} {J. Comput. Chem.}\ }\textbf {\bibinfo
  {volume} {17}},\ \bibinfo {pages} {1571} (\bibinfo {year}
  {1996})}\BibitemShut {NoStop}%
\bibitem [{\citenamefont {Schuchardt}\ \emph {et~al.}(2007)\citenamefont
  {Schuchardt}, \citenamefont {Didier}, \citenamefont {Elsethagen},
  \citenamefont {Sun}, \citenamefont {Gurumoorthi}, \citenamefont {Chase},
  \citenamefont {Li},\ and\ \citenamefont {Windus}}]{Schuchardt2007}%
  \BibitemOpen
  \bibfield  {author} {\bibinfo {author} {\bibfnamefont {K.~L.}\ \bibnamefont
  {Schuchardt}}, \bibinfo {author} {\bibfnamefont {B.~T.}\ \bibnamefont
  {Didier}}, \bibinfo {author} {\bibfnamefont {T.}~\bibnamefont {Elsethagen}},
  \bibinfo {author} {\bibfnamefont {L.}~\bibnamefont {Sun}}, \bibinfo {author}
  {\bibfnamefont {V.}~\bibnamefont {Gurumoorthi}}, \bibinfo {author}
  {\bibfnamefont {J.}~\bibnamefont {Chase}}, \bibinfo {author} {\bibfnamefont
  {J.}~\bibnamefont {Li}}, \ and\ \bibinfo {author} {\bibfnamefont {T.~L.}\
  \bibnamefont {Windus}},\ }\href {\doibase 10.1021/ci600510j} {\bibfield
  {journal} {\bibinfo  {journal} {{J.~Chem. Inf. Model.}}\ }\textbf {\bibinfo
  {volume} {47}},\ \bibinfo {pages} {1045} (\bibinfo {year}
  {2007})}\BibitemShut {NoStop}%
\bibitem [{\citenamefont {Sun}\ \emph {et~al.}(2017)\citenamefont {Sun},
  \citenamefont {Berkelbach}, \citenamefont {Blunt}, \citenamefont {Booth},
  \citenamefont {Guo}, \citenamefont {Li}, \citenamefont {Liu}, \citenamefont
  {Mcclain}, \citenamefont {Sayfutyarova}, \citenamefont {Sharma},
  \citenamefont {Wouters}, \citenamefont {Kin},\ and\ \citenamefont
  {Chan}}]{Sun2017}%
  \BibitemOpen
  \bibfield  {author} {\bibinfo {author} {\bibfnamefont {Q.}~\bibnamefont
  {Sun}}, \bibinfo {author} {\bibfnamefont {T.~C.}\ \bibnamefont {Berkelbach}},
  \bibinfo {author} {\bibfnamefont {N.~S.}\ \bibnamefont {Blunt}}, \bibinfo
  {author} {\bibfnamefont {G.~H.}\ \bibnamefont {Booth}}, \bibinfo {author}
  {\bibfnamefont {S.}~\bibnamefont {Guo}}, \bibinfo {author} {\bibfnamefont
  {Z.}~\bibnamefont {Li}}, \bibinfo {author} {\bibfnamefont {J.}~\bibnamefont
  {Liu}}, \bibinfo {author} {\bibfnamefont {J.~D.}\ \bibnamefont {Mcclain}},
  \bibinfo {author} {\bibfnamefont {E.~R.}\ \bibnamefont {Sayfutyarova}},
  \bibinfo {author} {\bibfnamefont {S.}~\bibnamefont {Sharma}}, \bibinfo
  {author} {\bibfnamefont {S.}~\bibnamefont {Wouters}}, \bibinfo {author}
  {\bibfnamefont {G.}~\bibnamefont {Kin}}, \ and\ \bibinfo {author}
  {\bibfnamefont {L.}~\bibnamefont {Chan}},\ }\href
  {https://arxiv.org/pdf/1701.08223.pdf} {\  (\bibinfo {year} {2017})},\
  \Eprint {http://arxiv.org/abs/1701.08223v2} {arXiv:1701.08223v2} \BibitemShut
  {NoStop}%
\bibitem [{\citenamefont {L{\"{o}}wdin}(1950)}]{Lowdin1950}%
  \BibitemOpen
  \bibfield  {author} {\bibinfo {author} {\bibfnamefont {P.~O.}\ \bibnamefont
  {L{\"{o}}wdin}},\ }\href {\doibase 10.1063/1.1747632} {\bibfield  {journal}
  {\bibinfo  {journal} {J. Chem. Phys.}\ }\textbf {\bibinfo {volume} {18}},\
  \bibinfo {pages} {365} (\bibinfo {year} {1950})}\BibitemShut {NoStop}%
\bibitem [{\citenamefont {Hinton}(2002)}]{Hinton2002}%
  \BibitemOpen
  \bibfield  {author} {\bibinfo {author} {\bibfnamefont {G.~E.}\ \bibnamefont
  {Hinton}},\ }\href {\doibase 10.1162/089976602760128018} {\bibfield
  {journal} {\bibinfo  {journal} {Neural Comput.}\ }\textbf {\bibinfo {volume}
  {14}},\ \bibinfo {pages} {1771} (\bibinfo {year} {2002})}\BibitemShut
  {NoStop}%
\bibitem [{\citenamefont {Handy}(1980)}]{Handy1980}%
  \BibitemOpen
  \bibfield  {author} {\bibinfo {author} {\bibfnamefont {N.}~\bibnamefont
  {Handy}},\ }\href {\doibase 10.1016/0009-2614(80)85158-X} {\bibfield
  {journal} {\bibinfo  {journal} {Chem. Phys. Lett.}\ }\textbf {\bibinfo
  {volume} {74}},\ \bibinfo {pages} {280} (\bibinfo {year} {1980})}\BibitemShut
  {NoStop}%
\bibitem [{\citenamefont {Knowles}\ and\ \citenamefont
  {Handy}(1984)}]{Knowles1984}%
  \BibitemOpen
  \bibfield  {author} {\bibinfo {author} {\bibfnamefont {P.}~\bibnamefont
  {Knowles}}\ and\ \bibinfo {author} {\bibfnamefont {N.}~\bibnamefont
  {Handy}},\ }\href {\doibase 10.1016/0009-2614(84)85513-X} {\bibfield
  {journal} {\bibinfo  {journal} {Chem. Phys. Lett.}\ }\textbf {\bibinfo
  {volume} {111}},\ \bibinfo {pages} {315} (\bibinfo {year}
  {1984})}\BibitemShut {NoStop}%
\bibitem [{\citenamefont {Sherrill}\ and\ \citenamefont
  {Schaefer}(1999)}]{Sherrill1999}%
  \BibitemOpen
  \bibfield  {author} {\bibinfo {author} {\bibfnamefont {C.~D.}\ \bibnamefont
  {Sherrill}}\ and\ \bibinfo {author} {\bibfnamefont {H.~F.}\ \bibnamefont
  {Schaefer}},\ }\enquote {\bibinfo {title} {{The Configuration Interaction
  Method: Advances in Highly Correlated Approaches}},}\ in\ \href {\doibase
  10.1016/S0065-3276(08)60532-8} {\emph {\bibinfo {booktitle} {Advances in
  Quantum Chemistry}}},\ Vol.~\bibinfo {volume} {34},\ \bibinfo {editor}
  {edited by\ \bibinfo {editor} {\bibfnamefont {P.~O.}\ \bibnamefont
  {L\"{o}wdin}}, \bibinfo {editor} {\bibfnamefont {J.~R.}\ \bibnamefont
  {Sabin}}, \bibinfo {editor} {\bibfnamefont {M.~C.}\ \bibnamefont {Zerner}}, \
  and\ \bibinfo {editor} {\bibfnamefont {E.}~\bibnamefont {Br\"{a}ndas}}}\
  (\bibinfo  {publisher} {Academic Press},\ \bibinfo {year} {1999})\ pp.\
  \bibinfo {pages} {143--269}\BibitemShut {NoStop}%
\bibitem [{\citenamefont {Pauncz}(1979)}]{Pauncz1979}%
  \BibitemOpen
  \bibfield  {author} {\bibinfo {author} {\bibfnamefont {R.}~\bibnamefont
  {Pauncz}},\ }\href {\doibase 10.1007/978-1-4684-8526-4} {\emph {\bibinfo
  {title} {Spin Eigenfunctions: Construction and Use}}}\ (\bibinfo  {publisher}
  {Springer US},\ \bibinfo {year} {1979})\BibitemShut {NoStop}%
\bibitem [{\citenamefont {Roos}(1994)}]{Roos1994}%
  \BibitemOpen
  \bibfield  {author} {\bibinfo {author} {\bibfnamefont {B.~O.}\ \bibnamefont
  {Roos}},\ }\href {\doibase 10.1007/978-3-642-57890-8} {\emph {\bibinfo
  {title} {Lecture Notes in Quantum Chemistry II: European Summer School in
  Quantum Chemistry}}},\ Vol.~\bibinfo {volume} {64}\ (\bibinfo  {publisher}
  {Springer-Verlag Berlin Heidelberg},\ \bibinfo {year} {1994})\BibitemShut
  {NoStop}%
\bibitem [{\citenamefont {Shankar}(1994)}]{Shankar1994}%
  \BibitemOpen
  \bibfield  {author} {\bibinfo {author} {\bibfnamefont {R.}~\bibnamefont
  {Shankar}},\ }\href {\doibase 10.1007/978-1-4757-0576-8} {\emph {\bibinfo
  {title} {Principles of Quantum Mechanics}}}\ (\bibinfo  {publisher} {Springer
  US},\ \bibinfo {year} {1994})\BibitemShut {NoStop}%
\bibitem [{\citenamefont {Davidson}(1975)}]{Davidson1975}%
  \BibitemOpen
  \bibfield  {author} {\bibinfo {author} {\bibfnamefont {E.~R.}\ \bibnamefont
  {Davidson}},\ }\href {\doibase 10.1016/0021-9991(75)90065-0} {\bibfield
  {journal} {\bibinfo  {journal} {J. Comput. Phys.}\ }\textbf {\bibinfo
  {volume} {17}},\ \bibinfo {pages} {87} (\bibinfo {year} {1975})}\BibitemShut
  {NoStop}%
\bibitem [{\citenamefont {Leininger}\ \emph {et~al.}(2001)\citenamefont
  {Leininger}, \citenamefont {Sherrill}, \citenamefont {Allen},\ and\
  \citenamefont {Schaefer}}]{Leininger2001}%
  \BibitemOpen
  \bibfield  {author} {\bibinfo {author} {\bibfnamefont {M.~L.}\ \bibnamefont
  {Leininger}}, \bibinfo {author} {\bibfnamefont {C.~D.}\ \bibnamefont
  {Sherrill}}, \bibinfo {author} {\bibfnamefont {W.~D.}\ \bibnamefont {Allen}},
  \ and\ \bibinfo {author} {\bibfnamefont {H.~F.}\ \bibnamefont {Schaefer}},\
  }\href {\doibase 10.1002/jcc.1111} {\bibfield  {journal} {\bibinfo  {journal}
  {J. Comput. Chem.}\ }\textbf {\bibinfo {volume} {22}},\ \bibinfo {pages}
  {1574} (\bibinfo {year} {2001})}\BibitemShut {NoStop}%
\bibitem [{\citenamefont {Li}\ and\ \citenamefont {Paldus}(1995)}]{Li1995}%
  \BibitemOpen
  \bibfield  {author} {\bibinfo {author} {\bibfnamefont {X.}~\bibnamefont
  {Li}}\ and\ \bibinfo {author} {\bibfnamefont {J.}~\bibnamefont {Paldus}},\
  }\href {\doibase 10.1063/1.469812} {\bibfield  {journal} {\bibinfo  {journal}
  {J. Chem. Phys.}\ }\textbf {\bibinfo {volume} {103}},\ \bibinfo {pages}
  {1024} (\bibinfo {year} {1995})}\BibitemShut {NoStop}%
\bibitem [{\citenamefont {Josef}\ and\ \citenamefont
  {Xiangzhu}(1999)}]{Paldus2007}%
  \BibitemOpen
  \bibfield  {author} {\bibinfo {author} {\bibfnamefont {P.}~\bibnamefont
  {Josef}}\ and\ \bibinfo {author} {\bibfnamefont {L.}~\bibnamefont
  {Xiangzhu}},\ }\enquote {\bibinfo {title} {{A Critical Assessment of Coupled
  Cluster Method in Quantum Chemistry}},}\ in\ \href {\doibase
  10.1002/9780470141694.ch1} {\emph {\bibinfo {booktitle} {Advances in Chemical
  Physics}}}\ (\bibinfo  {publisher} {John Wiley \& Sons, Inc.},\ \bibinfo
  {year} {1999})\ pp.\ \bibinfo {pages} {1--175}\BibitemShut {NoStop}%
\bibitem [{SM2()}]{SM2018}%
  \BibitemOpen
  \href@noop {} {}\bibinfo {note} {See Supplemental Material at [URL will be
  inserted by publisher] for additional information about the
  uncompressed/compressed molecular determinants.}\BibitemShut {Stop}%
\bibitem [{\citenamefont {Head-Gordon}(1996)}]{Gordon1996}%
  \BibitemOpen
  \bibfield  {author} {\bibinfo {author} {\bibfnamefont {M.}~\bibnamefont
  {Head-Gordon}},\ }\href {\doibase 10.1021/JP953665+} {\bibfield  {journal}
  {\bibinfo  {journal} {J. Phys. Chem.}\ }\textbf {\bibinfo {volume} {100}},\
  \bibinfo {pages} {13213} (\bibinfo {year} {1996})}\BibitemShut {NoStop}%
\bibitem [{\citenamefont {Lee}\ and\ \citenamefont {Taylor}(1989)}]{Lee2009}%
  \BibitemOpen
  \bibfield  {author} {\bibinfo {author} {\bibfnamefont {T.~J.}\ \bibnamefont
  {Lee}}\ and\ \bibinfo {author} {\bibfnamefont {P.~R.}\ \bibnamefont
  {Taylor}},\ }\href {\doibase 10.1002/qua.560360824} {\bibfield  {journal}
  {\bibinfo  {journal} {Int. J. Quantum Chem.}\ }\textbf {\bibinfo {volume}
  {36}},\ \bibinfo {pages} {199} (\bibinfo {year} {1989})}\BibitemShut
  {NoStop}%
\end{thebibliography}%

%%%%%%%%%% Merge with supplemental materials %%%%%%%%%%
\newpage
\widetext
\begin{center}
\textbf{\large Supplemental Material for ``Compression of Exact Wavefunctions with Restricted Boltzmann Machine Auto-Encoders''}
\end{center}
%%%%%%%%%% Merge with supplemental materials %%%%%%%%%%
%%%%%%%%%% Prefix a "S" to all equations, figures, tables and reset the counter %%%%%%%%%%
\setcounter{equation}{0}
\setcounter{figure}{0}
\setcounter{table}{0}
\makeatletter
\renewcommand{\theequation}{S\arabic{equation}}
\renewcommand{\thefigure}{S\arabic{figure}}
\renewcommand{\thetable}{S\arabic{table}}
\renewcommand{\bibnumfmt}[1]{[S#1]}
\renewcommand{\citenumfont}[1]{S#1}
%%%%%%%%%% Prefix a "S" to all equations, figures, tables and reset the counter %%%%%%%%%%

This supplemental material contains extra information about the uncompressed molecular configurations and the compressed ones.

\begin{table}[b!]

\caption{Bits per molecular configurations and number of configurations for FCI, spin-adapted (SA) CCSD(RHF) and the molecular RBM (mRBM). The basis set is STO-3G and the orbitals follow the L\"{o}wdin's symmetric orthogonalization scheme. \label{tab_S1}}\hypertarget{tab:table_S1}{}

\noindent \begin{centering}
\begin{threeparttable}
\begin{tabular}{ccccc}
\hline 
\hline
\noalign{\vskip0.12cm}
\textbf{Systems} & \textbf{BeH\textsubscript{2}} & \textbf{C\textsubscript{2}} & \textbf{N\textsubscript{2}} & \textbf{F\textsubscript{2}}\tabularnewline
\noalign{\vskip0.12cm}
\hline
\noalign{\vskip0.10cm}
\multicolumn{5}{c}{\textbf{Bits per molecular configurations}~\tnotex{tnote:bits_mconf}}\tabularnewline
\noalign{\vskip0.10cm}
FCI / SA CCSD(RHF) & 14 & 20 & 20 & 20\tabularnewline
\noalign{\vskip0.05cm}
mRBM & 10 & 18 & 15 & 7\tabularnewline
\hline
\noalign{\vskip0.10cm}
\multicolumn{5}{c}{\textbf{Spin-adapted (SA) molecular configurations}}\tabularnewline
\noalign{\vskip0.10cm}
(RHF $+$ S $+$ D)~\tnotex{tnote:sa_ccsd_confgs} & 91 & 325 & 253 & 55\tabularnewline
\hline 
\noalign{\vskip0.10cm}
\multicolumn{5}{c}{\textbf{Molecular configurations with $\langle\hat{S}_{z}\rangle=0$}~\tnotex{tnote:sz_operator}}\tabularnewline
\noalign{\vskip0.10cm}
FCI & 1,225 & 44,100 & 14,400 & 100\tabularnewline
\noalign{\vskip0.05cm}
Training Set & 212 & 6,688 & 1,880 & 48\tabularnewline
\noalign{\vskip0.05cm}
mRBM & 326 & 40,948 & 3,523 & 58\tabularnewline
%\hline 
%\noalign{\vskip0.10cm}
%\multicolumn{5}{c}{\textbf{Molecular configurations with $\langle\hat{S}_{z}\rangle\neq0$}~\tnotex{tnote:sz_operator}}\tabularnewline
%\noalign{\vskip0.10cm}
%mRBM & 17 & 0 & 1,768 & 2\tabularnewline
\hline
\hline
\noalign{\vskip0.08cm}
\end{tabular}
\begin{tablenotes}
      \item\label{tnote:bits_mconf}(See \textcolor{red}{\texttt{Assessing the Compression Sec.}} \emph{in paper} for details).
      \item\label{tnote:sa_ccsd_confgs}Molecular configurations; RHF~$=$~Restricted Hartree--Fock reference, S~$=$~singly excited and D~$=$~doubly excited. (See \textcolor{red}{\texttt{Results and Discussion Sec.}} \emph{in paper} for details).
      \item\label{tnote:sz_operator}$\langle\hat{S}_{z}\rangle$ is the expectation value of the $\hat{S}_{z}$ operator [the z-component of the spin operator]. For mRBM, $\langle\hat{S}_{z}\rangle$ is computed for the reconstructed configuration [decoding process].
    \end{tablenotes}
  \end{threeparttable}
\par\end{centering}
\vspace{0.5cm}
\noindent\noindent\hfil\rule{1.0\textwidth}{.4pt}\hfil
\end{table}

\indent Table~\protect\hyperlink{tab:table_S1}{S1} shows the number of bits per molecular configuration and the quantity of molecular configurations for FCI, mRBM and spin-adapted (SA) CCSD(RHF).

\indent For FCI and SA CCSD(RHF), the number of bits per configuration is associated to the amount of atomic orbitals of a system. On the other hand, for mRBM, this same number corresponds to the amount of hidden units of the trained RBM. 

\indent In its turn, the amount of hidden units is connected to the size of the training set, or, in other words, the RBM needs more hidden units to reconstruct training sets that include more molecular determinants. This can be seen by the ascending order of the training set---F\textsubscript{2} $<$ BeH\textsubscript{2} $<$ N\textsubscript{2} $<$ C\textsubscript{2}---which is equal as the ascending order of the bits per molecular determinants of the mRBM.

\indent Moving to the number of molecular configurations, Table \ref{tab_S1} exposes that SA CCSD(RHF) has few configurations when compared to FCI and mRBM for BeH\textsubscript{2}, C\textsubscript{2} and N\textsubscript{2}---since SA CCSD(RHF) only considers singly and doubly-excited singlet SA configurations. However, it is not the case for F\textsubscript{2}, when CCSD(RHF) becomes exact.

\indent For mRBM, the number of molecular configurations indicates that the molecular RBM recovers configurations which does not belong to the training set and also reduces the number of configurations which span the FCI wavefunction. This reduction of the configuration space is more pronounced for BeH\textsubscript{2}, N\textsubscript{2} and F\textsubscript{2}, and less pronounced for C\textsubscript{2}.

\end{document}